\documentclass{article}
\usepackage{amsmath,amsfonts}
\usepackage{algorithmic}
\usepackage{algorithm}
\usepackage{array}
\usepackage[caption=false,font=normalsize,labelfont=sf,textfont=sf]{subfig}
\usepackage{textcomp}
\usepackage{stfloats}
\usepackage{url}
\usepackage{verbatim}
\usepackage{graphicx}
\usepackage{cite}
\usepackage{multicol}
\usepackage{multirow}
\usepackage{hyperref}
\usepackage{authblk}
\begin{document}

\title{ADVENT: Attack/Anomaly Detection in VANETs}

\author[1,*]{Hamideh Baharlouei}
\author[2]{Tokunbo Makanju}
\author[1]{Nur Zincir-Heywood}
\affil[1]{Computer Science, Dalhousie University, Halifax, NS, Canada.}
\affil[2]{New York Institute of Technology - Vancouver Campus, Vancouver, BC, Canada.}
\affil[*]{Corresponding author(s). E-mail(s): hm729953@dal.ca}
\affil[ ]{Contributing author(s): amakanju@nyit.edu; zincir@dal.ca}

\maketitle

\begin{abstract}

In the domain of Vehicular Ad hoc Networks (VANETs), where the imperative of having a real-world malicious detector capable of detecting attacks in real-time and unveiling their perpetrators is crucial, our study introduces a system with this goal. This system is designed for real-time detection of malicious behavior, addressing the critical need to first identify the onset of attacks and subsequently the responsible actors. Prior work in this area have never addressed both requirements, which we believe are necessary for real world deployment, simultaneously. By seamlessly integrating statistical and machine learning techniques, the proposed system prioritizes simplicity and efficiency. It excels in swiftly detecting attack onsets with a remarkable F1-score of 99.66\%, subsequently identifying malicious vehicles with an average F1-score of approximately 97.85\%. Incorporating federated learning in both stages enhances privacy and improves the efficiency of malicious node detection, effectively reducing the false negative rate.  
\end{abstract}


\section{Introduction}
Vehicular Ad hoc Networks (VANETs) as a specific type of mobile ad-hoc networks (MANET) are commonly associated with transportation systems with three main channels: vehicle-to-vehicle (V2V), vehicle-to-roadside (V2R), and vehicle-to-infrastructure (V2I), collectively known as V2X (vehicle-to-everything). VANET communications provide real-time information to tackle delayed driver responses. This enables immediate control over vehicle functions like brakes, acceleration, and steering. It offers advantages such as contributing to traffic safety by delivering precise information directly to drivers. However, the dynamic nature of VANETs, marked by constantly changing network topologies, varying vehicle speeds, and differences in the density of V2X communications, introduces new challenges and vulnerabilities that must be addressed \cite{garip}. These vulnerabilities can be exploited to launch various types of attacks, which could result in various issues such as accidents and traffic congestion. Thus, ensuring the security of VANETs is of great significance due to the potential risks to human lives, property, and economic activities. This underscores the need to prioritize the development of robust information system security tools and mechanisms capable of not only detecting but also effectively mitigating these attacks. Taking proactive measures is essential to ensure the integrity and safety of VANETs in the face of the evolving cybersecurity threats.\cite{garip,8888176,a2020misbehavior}

Unfortunately, conducting research in these fields through the deployment and testing of actual VANETs often entails significant expense and labour, making it impractical in many instances. Therefore, simulation tools are widely employed to evaluate strategies before the actual implementation of various algorithms and methodologies \cite{martinez2011survey}. In this study, we utilized datasets generated via simulations, as outlined in \cite{ours}. These simulations take into account three main factors to ensure the reliability of the results for real-world applicability. The first factor is the grid pattern. To replicate scenarios in our simulations, we leverage real geographical maps and topological data from multiple cities with different grid patterns. These different grid patterns can impact vehicle behavior, movements, and communication. This approach provides a better understanding of how the proposed system could work in the real-world conditions as it has been tested in city road networks with different grid patterns.  The second factor considered is the integration of realistic communication settings within the simulations. This directly influences how data packets are transmitted and received, aligning the simulations more closely with real-world conditions. The third and final factor revolves around the mobility patterns of vehicles. This encompasses various aspects, including vehicle movement patterns, the number of vehicles in each scenario, their entry into the simulation (including time intervals), their trajectories, and their interactions within the network. In this research, these aspects are designed to mirror real-world vehicle behaviors and movements, thereby also enhancing the practicality of the simulation results \cite{martinez2013computer,ours}. 

DDoS attacks pose a significant threat to the reliability and functionality of VANETs, as these networks heavily rely on the exchange of real-time information among vehicles and infrastructure to enhance road safety and efficiency \cite{hasrouny2017vanet, sakiz2017survey}. By flooding the network with an overwhelming volume of malicious traffic or by exploiting vulnerabilities in the network infrastructure, attackers can exhaust network resources, degrade the quality of service, and impede the transmission of vital safety messages. This can lead to delays, loss of critical information, and compromised situational awareness among connected vehicles, ultimately endangering the safety of drivers and passengers. Addressing DDoS attacks in VANETs is thus crucial, as it ensures the network's integrity, reliability, and uninterrupted communication, all essential for safe and efficient transportation \cite{de2014detecccao}. 

In this paper, we propose a real-time malicious behavior detector which operates in three distinct steps that include, data preprocessing, the detection of the onset of an attack and the identification of the malicious nodes involved. It combines both Machine Learning (ML) and statistical techniques to leverage the advantages of each method. Furthermore, it ensures crucial data privacy, a significant concern in all ML-based approaches, especially when training with data collected from various nodes or locations and sharing it with other nodes in the network. This privacy is achieved through the integration of federated learning \cite{FLXGBOOST}. Federated learning is seamlessly integrated into both the machine learning and statistical approaches for identifying attack onsets and malicious nodes. This integration allows us to avoid transmitting raw data to other nodes in the network, instead, sharing only trees and lists. Additionally, the inclusion of Federated Learning significantly improves efficiency and notably reduces the False Negative Rate (FNR), particularly in the malicious node detection phase, while preserving crucial privacy. It's worth emphasizing that while ML-based approaches offer advantages, they come with certain challenges as well. The effectiveness of ML algorithms relies heavily on the quality of the training data, preprocessing and feature selection. These challenges, if not properly addressed, can lead to faulty and unnecessarily complex models.

To address these challenges, we have introduced an innovative preprocessing method. This method generates training data that captures the number of packets received by each node in every second within the VANET network. By adopting this approach, we gain a more comprehensive understanding of communication patterns, potential attack scenarios, and facilitate real-time operation. In particular, this method reduces the dataset's size and processing time, contributing to a more efficient exploration of network dynamics. Consequently, it enhances our capabilities for effective and efficient attack detection in VANETs. Additionally, we are able to determine a threshold for the ratio of benign to malicious samples required in data for local model training. Our proposed system employs this threshold value and Synthetic Minority Oversampling Technique (SMOTE) \cite{smote} to achieve the desired ratio in all local training samples. This has proven instrumental in obtaining improved results and minimizing the False Negative Rate (FNR).
\noindent{\bf In summary, this research introduces five novel contributions:}

\begin{enumerate}
\item{Designing and developing a comprehensive three-tier real-time system that effectively detects the onset of an attack and identifies the malicious nodes responsible.} 
\item{Implementing a Federated Learning based approach for enhaning privacy, while simultaneously leveraging the collective experience of all nodes to boost the efficiency and reduce the False Negative Rate (FNR).} 
\item{Prioritizing simplicity and efficiency through the feature selection process, which combines domain expertise with innovative feature engineering techniques. This enables the selection of the most informative features, increasing the ability of the proposed system to detect attacks in VANETs.}
\item{Bolstering the robustness of the proposed system via datasets that are generated in realistic VANET simulations, with a focus on geographical and traffic diversity. Thus, unlike prior work \cite{MTH}, datasets employed represent the VANET dynamics more accurately. The use of various geographical maps and simulation factors ensures the reliability of data, which serves as a strong foundation for training and evaluating the proposed system \cite{ours,ours2}}
\item{Addressing the challenge of imbalanced data and the scarcity of malicious samples in local dataset via threshold identified for mitigation. Subsequently, the Synthetic Minority Over-sampling Technique (SMOTE) is implemented based on this threshold. This strategy aims to balance the dataset by generating synthetic samples, particularly focusing on the minority class, and therefore, contributing to more robust and representative training data for the proposed system.}
\end{enumerate}

The rest of this paper is organized as follows: Section \ref{sec:prior-work} discusses previous works. Section \ref{sec:methodology} describes the components of the proposed system and how they work. Simulations and results are presented in Section \ref{sec:results}. Finally, conclusions are drawn and future research directions are discussed in Section \ref{sec:con}.

\section{Related work} \label{sec:prior-work}

In this section, a comprehensive overview of previous studies in the domain of Vehicular Ad hoc Networks (VANETs) is presented, with a specific focus on three key aspects: the general landscape of VANETs, the application of machine learning for attack detection in VANETs, and the exploration of Federated Learning within the context of VANETs. 

\subsection{VANETs}
Cavalcanti et al. \cite{Cavalcanti} extensively reviewed a decade of VANET research (2007-2016) in their analysis of 283 scholarly works. Their findings highlighted the enduring prevalence of simulation as the primary validation tool for solutions in MANETs. Over 75\% of papers consistently utilized simulation, a trend that has persisted since 2000-2005. The study also noted a shift from Network Simulator 2 (NS-2) to OMNET++, with a 19.5\% decrease in NS-2 usage and a more than twofold increase in OMNET++ adoption, from 5\% to 12\%. Moreover, Cavalcanti et al. explored choices of simulation tools for vehicular networks, identifying SUMO as the most widely used (31.2\%), followed by VanetMobiSim at 11.8\%. However, a significant 42.4\% of simulation-based papers did not specify the mobility tool used, indicating the diverse and evolving landscape of VANET research methodologies.

\subsection{Using Statistical Techniques for Malicious Node Detection}

Valentini et al. \cite{MDASTI} proposed MDASTI (Attacks Detection Mechanism for Intelligent Transport System) as a host-based attack detection mechanism for intelligent transport systems, focusing on ARP protocol data. Their approach utilized the MAD statistic for identifying potentially malicious nodes, particularly in the context of Sao Paolo, where they calculated a threshold for the number of normal packets received from each neighbor and made a list of malicious nodes. 

Baharlouei et al. \cite{ours} made contributions to the field by leveraging Malicious Activity Detection (MAD) to characterize Distributed Denial of Service (DDoS) attacks. Their research delved into the intricacies of DDoS attacks across a spectrum of urban scenarios, considering factors such as packet transmission ranges and the number of vehicles involved in different scenarios. Their findings show the critical importance of conducting realistic simulations, particularly when evaluating Intrusion Detection Systems (IDSs). In the realm of cybersecurity, where precision and reliability are paramount, the need for realistic simulations cannot be overstated.
Valentini et al. \cite{MDASTI} and Baharlouei et al. \cite{ours} have made notable contributions to vehicular network security by preparing datasets and detecting malicious nodes. But their work did not encompass the critical aspects of attack onset detection and privacy preservation. These are significant facets that this work aims to address and enhance, thereby providing a more comprehensive and robust solution for vehicular network security.  

\subsection{Using Machine learning for Attack Detection}
Alheeti et al. \cite{20} introduced an intelligent intrusion detection system (IDS) that leveraged back-propagation neural networks for identifying Denial of Service (DoS) attacks. Their approach, which relied on the Kyoto 2006+ dataset, incorporated ensemble learning and feature selection to achieve a commendable balance between high detection rates and low computational costs.
%%%%%%%%

Rani et al. \cite{RANI}, on the other hand, proposed an intelligent transport system tailored to Internet of Vehicles (IoV)-based vehicular network traffic. Their system harnessed tree-based models, including Decision Tree (DT), Random Forest (RF), Extra Tree (ET), and XGBoost, demonstrating substantial detection accuracy and computational efficiency on the "CIC-IDS2017" dataset by employing ensemble learning and essential feature selection techniques.

Aloqaily et al. \cite{23} introduced an IDS for IoV and connected vehicles, relying on deep belief networks (DBN) and decision tree (DT) algorithms. Their system exhibited impressive accuracy, as validated using the NSL-KDD dataset.

Gao et al. \cite{24} addressed the challenge of distributed Denial of Service (DDoS) network intrusion detection. Their system was composed of two key modules: a real-time network traffic collection module and a network traffic detection module. The latter employed a classification algorithm based on Random Forest (RF) and was evaluated and compared using datasets that included NSL-KDD and UNSW-NB15.

Additionally, Injadat et al. \cite{28} devised a multi-stage, optimized machine learning-based IDS specifically tailored for network attack detection. Their performance evaluations were conducted on the CICIDS2017 and UNSW-NB15 datasets, where they leveraged hyper-parameter optimization (HPO) techniques to fine-tune machine learning models.

Li et al. \cite{9013892}conducted a thorough survey on Controller Area Networks (CAN) and Autonomous Vehicle (AV) networks' vulnerabilities, culminating in an Intrusion Detection System (IDS) designed for both contexts. They harnessed four tree-based supervised learning algorithms and integrated a stacking ensemble model to enhance intrusion detection accuracy while optimizing computational efficiency. Evaluation using public datasets demonstrated the IDS's high accuracy in safeguarding CAN and AV networks, emphasizing the efficacy of tree-based supervised learning in Internet of Vehicles and autonomous driving security.
Previous studies, such as those by Alheeti et al. \cite{20}, Rani et al. \cite{RANI}, Aloqaily et al. \cite{23}, Gao et al. \cite{24}, Injadat et al. \cite{28} and Li et al. \cite{9013892}, have explored various intrusion detection methods for vehicular networks. These studies encompassed a wide array of techniques, from back-propagation neural networks to tree-based models and used different datasets for evaluation. The proposed work not only builds upon this prior research but also fills a critical gap in vehicular network security by focusing on attack onset detection, malicious node identification, and robust data privacy techniques, all while maintaining model efficiency and effectiveness and providing evaluations based on data from VANETS.

In a recent work, Li et al. \cite{MTH}introduced a robust multi-tiered hybrid Intrusion Detection System (MTH-IDS) for IoV environments. This IDS effectively safeguards intra-vehicle and external networks from diverse attack types, both known and unknown, operating in real-time scenarios. MTH-IDS involves data pre-processing, feature engineering, and four tiers of learning models, addressing various aspects of known and unknown attack detection.
Evaluation using well-established datasets (CAN-intrusion-dataset\cite{Seo_2018} and CICIDS2017-dataset\cite{Sharafaldin2018TowardGA}) demonstrated MTH-IDS's remarkable performance in detecting unknown attacks, maintaining a high detection rate with a low false alarm rate. This research underscores MTH-IDS's potential as a robust and versatile intrusion detection system for Internet of Vehicles environments.

Baharlouei et al. \cite{ours2} introduced an approach for detecting malicious behavior in VANETs, covering both attack onset and identification of malicious nodes. They utilized a real VANET dataset, incorporating realistic simulation factors and introducing a novel data preprocessing method. The authors employed four tree-based models to construct an ensemble model, optimizing parameters with BO-TPE (Bayesian Optimization with Tree-structured Parzen Estimator). They implemented a malicious node detection mechanism using the MAD method, achieving a 100\% Detection Rate in the Attack Onset detection step and an average of 90\% in the Malicious Node detection step.
%%%%%%%%%%%%%%%%%%%%%%%%%%
\subsection{Federated Learning}
Cheng et al. \cite{SecurelosselessXG} address the critical concern of user privacy protection in machine learning and propose SecureBoost. This innovative federated learning system achieves privacy preservation by conducting entity alignment through a privacy-preserving protocol and subsequently constructing boosting trees across multiple parties using a meticulously designed encryption strategy. SecureBoost operates on vertically partitioned datasets, where multiple parties collaborate in the learning process while maintaining distinct feature sets.

Ma et al. \cite{FLXGBOOST}address the privacy challenges associated with decentralized datasets and leverages the strength of extreme Gradient Boosting (XGBoost) in handling tabular data within the context of federated learning (FL). Notably, they tackle a specific aspect of FL, horizontal federated XGBoost, leading to heightened per-node communication frequencies and substantial privacy concerns. To mitigate these issues, Ma et al. introduce an innovative system that removes the dependency on gradient sharing. Instead, they make the learning rates of aggregated tree ensembles learnable, simultaneously enhancing privacy and communication efficiency. Through extensive evaluations on diverse datasets for classification and regression tasks, their approach demonstrates performance comparable to state-of-the-art methods. Moreover, it significantly enhances communication efficiency by reducing both communication rounds and overhead.

\subsection{Drawbacks of Prior Work}
While \cite{MTH,9013892} represents the state-of-the-art, 
Yang et al. dedicated their efforts to the critical task of attack detection, although they did not specifically address malicious node detection. Their research involved the utilization of two datasets, one for internal vehicle networks and another for external network traffic. However, it is worth noting that their work did not include simulations of vehicular network traffic.
Valentini et al. addressed malicious node detection in \cite{MDASTI} but overlooked the onset of an attack, using only one dataset from Sao Paolo. None of these papers incorporated federated learning in VANET security or attack detection. Privacy concerns were insufficiently addressed, especially during training, where entire datasets were employed without considering privacy implications. In the real world, using actual vehicle datasets for training raises significant privacy issues. Moreover, none discussed incorporating other vehicles' experiences in malicious node detection. By leveraging federated learning, we offer the benefits of shared experiences without compromising privacy.

While they also utilized Synthetic Minority Over-sampling Technique (SMOTE)\cite{smote}, it's noteworthy that our approach distinguishes itself by incorporating a specific threshold and a method for applying SMOTE based on this identified threshold. By establishing a relationship between the required number of samples for efficient model training, we have implemented a targeted and informed use of SMOTE. This methodological nuance enhances the adaptability and effectiveness of our data augmentation strategy, addressing imbalanced datasets more precisely for improved model performance.

In \cite{ours},  Baharlouei et al. focused on VANET simulations and detecting malicious nodes. However, they did not study the detection of the onset of an attack either. In \cite{ours2}, Baharlouei et al. delved into real-time malicious behavior detection, identifying the onset of attacks and malicious nodes. However, two significant unresolved issues emerged. Firstly, they did not address privacy concerns when utilizing raw data from all nodes for model training, a critical matter especially when dealing with data from real word cyber-phyiscal systems that might not be owned or controlled by the same entity. Secondly, there was a notable variation in the False Negative Rate (FNR) during malicious node detection due to differences in how each node experienced the attack. Based on their position in the grid and the length of time they remained in the grid during the attack, some nodes were only partially impacted by the effects of the attack. This diversity posed challenges in determining the optimal threshold and identifying malicious nodes. In this context, the development of a method allowing the simultaneous utilization of all nodes' experiences could mitigate these issues arising from topological and geographical disparities. Additionally, leveraging federated learning for decentralized model training proved instrumental in streamlining our approach. This approach reduced the number of models to just one in the attack onset detection step and eliminated the need for an optimization step. Both of these adjustments effectively reduced complexity and enhanced overall performance, enabling the development of a real-time method.
\begin{table*}[!t]
\caption{Comparative Analysis of Contemporary Intrusion Detection Methods for Vehicular Ad-Hoc Networks (VANETs)}
\label{tab:related}
\centering
\begin{tabular}{|c||c||c||c||c||c||c||c||c||c||c||c|}
\hline
Paper Authors & VD &AD
& DOA &MND & FE & SL&FL  & DP & AM & OT & ID\\

\hline
Valentini et al.\cite{MDASTI} &* & & & * & * &  &  &&&&\\
\hline
Yang et al. \cite{9013892}  & & *  &  &   &*   &  & &&&&\\
\hline
Yang et al. \cite{MTH}  &&  * &  &  & * & * & &&&*&*\\
\hline
Cheng et al. \cite{SecurelosselessXG}& & &  & & &  & *&*&*&&\\
\hline
Ma et al. \cite{FLXGBOOST}&  & &  &   &  &   & *&*&*&&\\
\hline
Baharlouei et al.\cite{ours2} &*& *  & * &  * & *& *& &&&*&\\
\hline
Proposed System &*& *  & * &  * & * &  * & *&*&*&*&*\\

\hline
\end{tabular}
\end{table*}

In Table \ref{tab:related}, we provide a comprehensive summary that compares the work presented in this paper to some of the most relevant prior works mentioned earlier. The comparison is carried out using what we consider the key components for evaluating research into Intrusion Detection System (IDS) for Vehicular Ad Hoc Networks (VANETs) and the components the system should have to be ready for real world deployment. The aspects we considered including, the utilization of VANET datasets(VD), attack detection(AD), capability in detecting the onset of attacks(DOA), identifying malicious nodes(MND), incorporation of feature engineering(FE), application of supervised learning(SL), integration of Federated Learning(FL), provision of data privacy measures(DP), utilization of aggregation methods(AM), optimization techniques(OT) employed, and the extent to which each method addresses issues related to imbalanced data(ID). This comparative analysis provides a holistic overview of the strengths and limitations of previous approaches in the field.
In summary our research presents a novel feature representation method and a multi-layered real-time approach capable of detecting the initiation of a DDoS attack and identifying the responsible malicious nodes. We prioritize data privacy for all nodes by leveraging federated learning techniques, sharing only trees instead of raw data during attack onset detection and lists during malicious node detection. Additionally, we conduct evaluations using four distinct vehicular network datasets that represent real-world scenarios, providing a comprehensive assessment of our proposed approach.

\section{Methodology/System Overview}

\label{sec:methodology}
\begin{figure}[!t]
\centering
\includegraphics[width=5in]{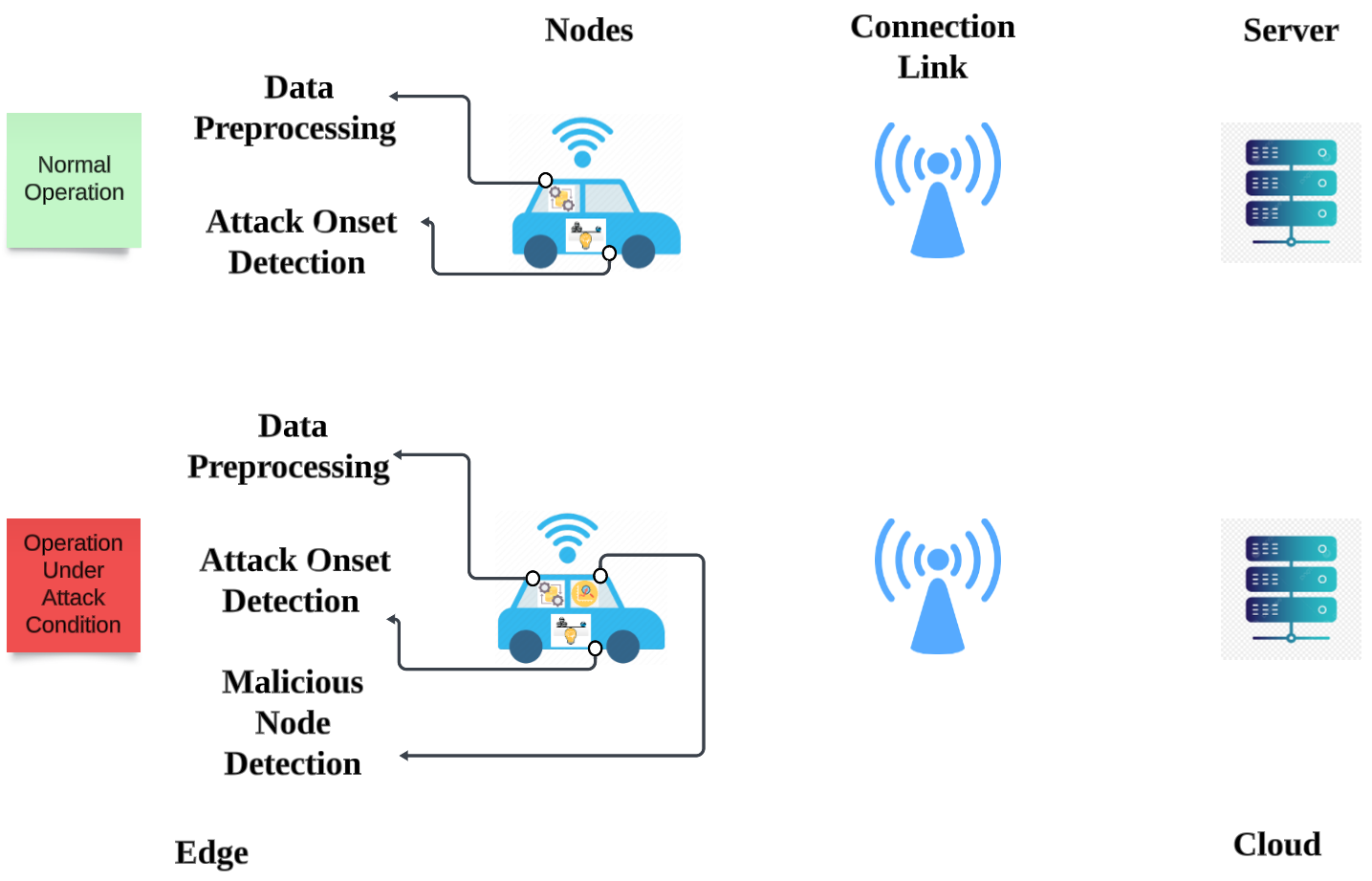}
\caption{Proposed system: Attack/Anomaly Detection in Vehicular NeTworks (ADVENT)}
\label{frame}
\end{figure}

In this paper, a multi-layered real-time malicious behavior detection system is proposed. The system aims to identify the onset of DDoS attacks and, subsequently, the malicious vehicles involved. The system architecture, illustrated in Figure \ref{frame}, encapsulates the essential components that collaboratively contribute to achieving our objectives. In the following sections, we delve into the details of these components and elucidate how the system operates under various conditions.

\subsection{Components of the System}
The proposed system comprises three main components: nodes (vehicles), server in the cloud, and communication links connecting them. All of these are described in this section.

\subsubsection{Nodes (Vehicles)}
Upon joining the network, each vehicle initiates data processing using the proposed system, creating its labeled dataset from observed data streams. During regular operation, there is no need for the Malicious Node Detection module. However, vehicles periodically train new attack onset detection models, sharing them with the server in the cloud at predetermined intervals to maintain an updated model across the network. Additionally, all nodes continue to engage in regular V2V communication to exchange information. 

\subsubsection {Central Server in the Cloud}
The central server plays a crucial role in the system's initialization process by acquiring initial trees from simulated or real-world data during the evaluation phase. Upon joining the network, each vehicle receives the latest model from the server, establishing a synchronized starting point for all nodes. Subsequently, as nodes contribute their local models and share them with the server, the model aggregation process is initiated, detailed in Section \ref{A_ONSET}. The server then facilitates the aggregation of local malicious vehicles lists and furnishes the aggregated list to each vehicle in the cloud. This serves as a pivotal element in the system's overall functionality.

\subsubsection{Communication Links}
Communication links serve as a vital conduit for efficient data transfer, with vehicles actively contributing to the central server's repository of the latest models. This bidirectional communication ensures that both the server and vehicles maintain current and synchronized information. 

Communication links, providing the connection to the server, incorporates components that facilitate efficient data transfer. We anticipate leveraging cellular networks for reliable data transmission from vehicles to the central server. Alternatively, low Earth orbit satellite communications, or radio technologies are available for connections between vehicles and the server. The choice of the communication technology (links) depends on the specific requirements and environmental conditions.

The potential inclusion of Road Side Units (RSUs) is an additional consideration for enhancing communication, but the proposed system is designed to operate seamlessly with or without them. One of the key strengths of the proposed system is its adaptability to various communication links and methods. It does not rely on any particular type of communication technology, offering the flexibility to be applied in diverse scenarios. Given that the proposed system does not involve the sharing of sensitive data that could be reconstructed by attackers, conventional communication security methods are sufficient for safeguarding the system.

\subsubsection{Cold Start Phase}
For the 'cold start' phase the ADVENT system leverages a combination of initializations for the central server and immediate model distribution for new nodes joining the network. By incorporating these strategies, we ensure a seamless integration process and maintain synchronization within the VANET environment.
The central server, residing in the cloud, undergoes a crucial initialization process at the onset of the Cold Start. To bootstrap its functionality, the server taps into the wealth of knowledge derived from simulations or real-world data as carried out in our evaluations. These initial simulations equip the server with fundamental insights into potential attack scenarios, communication patterns, and network dynamics.

When a new node joins the network, a key challenge lies in swiftly integrating it into the existing ecosystem. Our system addresses this challenge by adopting an immediate model distribution approach. Upon successful authentication with the central server, a new node promptly receives the latest model, ensuring that it is synchronized with the current state of the network. This immediate integration process allows the newcomer to actively contribute to the collaborative detection efforts without delay.

By leveraging a combination of simulation-derived initialization for the central server and immediate model distribution for new nodes, our system guarantees a seamless integration process. This integration is not only efficient but also critical for maintaining synchronization within the VANET environment. The synchronized state serves as the linchpin for effective collaboration among nodes and the central server.

The Cold Start phase lays a robust foundation for the system's continuous operation. It functions as a catalyst, enabling the system to adapt to varying network conditions, detect malicious behavior in real-time, and respond dynamically to changes within the VANET environment. This strategic initialization approach ensures that the system is well-prepared to navigate the intricacies of VANETs and provide reliable security solutions throughout its operational lifecycle.

\subsubsection{Data Processing}
Vehicles actively process their received packets using our proposed methods. Thus creating a continuous stream of data that they use in monitoring their communications or creating new detection models to be shared with the server their. This continuous operation forms the basis for ongoing model refinement and adaptation to the evolving network conditions.

\subsection{System's Regular Operation}

Each vehicle, upon joining the network, undergoes a standardized initialization process thorugh receving the latest model from the central server. The communication link facilitates the exchange of information, allowing vehicles to actively contribute to the central server's model repository. Thus, under regular operation, when no attack is underway, the components collaborate to monitor and secure the network.

Moreover, during regular operation, each vehicle processes its data stream using the proposed system. Thus, each vehicle monitors its own observations, which in return labels its own data based on what the vehicle observes. Vehicles also train new attack onset detection models, ensuring that both the server and individual vehicles maintain up-to-date models. 

This synchronized operation under normal conditions enables the system to proactively adapt to changing network dynamics, fostering a resilient and efficient environment. The methodology prioritizes privacy by leveraging each vehicle's data for individual model training. Each node independently constructs its model using local data without sharing this data with server, reinforcing privacy measures within the system. The regular updates contribute to the accuracy and relevance of the overall system.

\subsubsection{Model Building: From the Perspective of the Server in the Cloud}

In this system the XGBoost model is used, it belongs to the family of boosting algorithms, which sequentially train a series of weak learners, typically decision trees, to collectively form a robust predictive model \cite{61}.Our approach utilizes this tree-based model to identify the onset of an attack. In this paper, extend the approach based on the methodology presented in \cite{FLXGBOOST} by leveraging FedXGBllr to simultaneously update our hyperparameters and enhance privacy. The model is executed locally by each vehicle, resulting in the creation of individual trees and weights. Subsequently, these vehicles establish connections with the server to update the tree ensemble and weight parameters for subsequent rounds. Once the server distributes the updated trees, each vehicle becomes capable of detecting the onset of an attack. Upon detection, the Malicious Node Detection component is activated to identify potential attackers.

Under normal/regular operation, nodes actively participate in regular model building activities, contributing to the collaborative effort of maintaining an up-to-date model across the network. In this paper, four potential approaches for model updates are proposed. The first approach invloves using one model for a static interval, initiating the model preparation phase based on the chosen interval. While this approach allows for flexibility in adjusting the interval based on network conditions, it could cause repeating the model preparation phase even when the network is stable. this could then lead to unnecessary overhead.

The second approach involves counting the number of new nodes during the authentication phase and setting a threshold. If the number of newly joined nodes surpasses this threshold, the model preparation phase is triggered. This method leverages the data from newly joined nodes for model building and avoids unnecessary processes and overhead when the network is stable. However, determining the appropriate threshold becomes crucial, as it needs to balance the real network requirements.

The third approach involves renewing models based on the number of detected attacks in the network. When a pre-determined threshold of attack onsets are detected, the models are renewed, allowing nodes to use the updated models for detecting subsequent attack onsets. While this ensures the system takes advantage of all up-to-date data, determining the suitable threshold for the number of attacks becomes essential to optimize the benefits.

The fourth approach proposes a comprehensive solution by combining different strategies, creating a balanced and customized response to the network's specific requirements. In this approach, each incident is assigned a weight, and a threshold is defined for the cumulative weight. If the combined weight surpasses the established threshold, the system triggers the model renewal process, ensuring the preparation of an up-to-date model. However, in the absence of ongoing attacks and if no new vehicles meet the server's new arrival threshold, nodes can continue operating with the existing model, avoiding unnecessary updates. This approach optimizes efficiency by activating the model update process only when necessary, thereby aligning with the system's dynamic adaptation to varying network conditions.

\subsection{System's Operation under Attack} 

When a node detects the onset of an attack, it initiates the malicious node detection component. This componenet generates a suspicious actor list based on the data it observes and shares the list with the server. Upon receiving the initial suspicious list, the attack operation phase commences at the server. The first step in this phase involves determining the number of vehicles reporting an attack onset, and enabling the server to assess whether it might be a false positive. If the server confirms a real attack, it proceeds to the next phase. In this phase, the server utilizes the previous list of malicious nodes and the newly received suspicious nodes to prepare the final list, which is then broadcasted to all connected nodes. Simultaneously, the server considers the process of updating the model based on the new data received, especially if this is the first attack onset detection after the cold start process.

\subsubsection{ Malicious Node Detection}
Upon detecting an attack onset, vehicles promptly activate the malicious node detection module. Through communication with the server, multiple reports from different vehicles are aggregated, enabling the server to validate and confirm the attack. To prepare the final malicious list, the server considers all suspicious lists received from nodes, implementing a threshold for the number of reports for each suspicious node to prevent false positives. Subsequently, the server aggregates and distributes the malicious node list to all nodes. This collaborative approach empowers the network to collectively respond to confirmed attacks, effectively minimizing the impact of false positives. The distributed list ensures that every node remains informed about potential threats, promoting a unified response to enhance overall network security.

In this step, two different approaches are considered for updating the malicious list. The first approach is a stateless approach. This includes the server creating the list each time and sending the final list to the nodes without retaining it for future use. This approach is valuable considering the highly changeable nature of VANETs. However, if an attacker remains silent for a period and then resumes the attack, both the server and nodes would forget about the previous state, necessitating a new detection process.

On the other hand, the second approach is a stateful approach. this includes the server keeping the last list and setting a timer for each malicious node. If the timer expires and there is no new report for the malicious node, the server removes it from the next list. However, if the malicious node is reported again, the server resets the timer and retains the node in the list. This approach allows the system to maintain a memory of previous experiences, avoiding the need to repeat the entire detection process for a previously identified attacker. Nevertheless, keeping the timers up-to-date, setting them, and deciding on the timer's duration entail additional overhead for the system.

\section{Evaluations}
\label{sec:results}
In order to assess the performance and effectiveness of the proposed system, we conducted a series of extensive experiments encompassing various attack scenarios, which resulted in the creation of diverse datasets. Detailed descriptions of these experiments are provided in the following subsections.

\subsection{Simulated Dataset Creation}
As previously mentioned, we utilized the Instant Veins5.1-i2 simulation environment \cite{VIENS} to generate vehicular network data. Veins comprises OMNeT++ 5.6.2 \cite{OMNET} and SUMO 1.8.0 \cite{SUMO}, offering a comprehensive simulation suite. Our simulations include four urban cities: Sao Paulo (Brazil), Manhattan (New York, USA), Valencia (Spain), and Teruel (Spain).

We conducted six distinct simulation runs in each city, varying the percentage of attackers from 5\% to 30\% in increments of 500 basis points. Each simulation had a duration of 3600 seconds, equivalent to 24 real-world hours. Within this timeframe, vehicles entered and exited the simulation after traversing multiple edges on their routes to destinations. The number of edges each vehicle traversed and the specific edges they crossed were determined by the routes assigned within Sumo maps. The time intervals for adding vehicles were adjusted based on the number of vehicles and the simulation's duration. With 360 vehicles and a 3600-second simulation, the interval was set to 9.5 seconds to allow sufficient time for the last vehicles to complete their trips. The initial ten vehicles were introduced within the first ten seconds in all scenarios, resulting in a range of 10 to 30 vehicles being present in the grid during the simulations at any given time.

Throughout each simulation, we executed six attacks, each spaced 600 seconds apart. Each attack had a duration of 25 simulation seconds, equivalent to 25 real-world minutes. When an attacker was present in the simulation, it engaged in attacks against other vehicles for the entire 25-second period or for the duration of its presence. At all other times, malicious nodes behaved normally while within the simulation. We collected data from all vehicles during the simulation, subsequently analyzing this data to evaluate the method's performance and efficiency under varying attack scenarios and geographical conditions. For clarity and easy reference, the simulation configuration details are presented in Table \ref{tab:table2}.
 
\begin{table}[!t]
\caption{Simulation Parameters}\label{tab:table2}
\centering
\begin{tabular}{|c||c|}
\hline
Simulation Time (Seconds)      & 3600                          \\
Route Scenario                     & Realistic urban                    \\ 
Mobility Model                     & SUMO                \\
Propagation Model    &    Two-Ray Ground \\
Physical and MAC   Layer           & IEEE 802.11p                       \\
Routing Protocol                   & AODV                               \\
Transmission Rate                  & 6Mbps BW10MHz              \\
Radio Coverage (Meters)                     & 400                   \\ 
Transmission Power (dBm)                & 33.8 -- 60.0                \\
Transport Protocol & UDP  \\
Total Vehicles                     & 360                          \\
Total Serving   Vehicles           & 10 - 30                           \\
Total Malicious   Vehicles         & 5\% - 30\%                         \\
Attack Duration Time (Seconds)              & 25                    \\
Number Of Attacks                           & 6 \\
\hline

\end{tabular}
\end{table}

\begin{table}[!t]
\caption{Summary of the Datasets Used}\label{tab:table3}
\centering
\begin{tabular}{|c||c||c||c|}
\hline
Dataset & Total \# of
& \# of benign 
& \# of  malicious\\
&  Samples &  Samples &  Samples\\
\hline
Manhattan-5\% & 4,676,165 &    4,039,228
&     636,937\\
\hline
Manhattan-10\% & 5,083,855  &     3,995,742 & 1,088,113\\
\hline
Manhattan-15\% & 5,794,656 &    3,908,557
&   1,886,099\\
\hline
Manhattan-20\% & 6,253,454 &     3,856,242
&    2,397,212\\
\hline
Manhattan-25\% & 6,622,686 &     3,856,103
&   2,766,583\\
\hline
Manhattan-30\% & 7,248,367 &    3,735,463
&  3,512,904\\
\hline
Sao Paolo-5\% & 12,986,664 &    7,054,112
&   5,932,552\\
\hline
Sao Paolo-10\% & 1,336,3751 &  10,880,979
&   2,482,772\\
\hline
Sao Paolo-15\% & 14,184,846  &   10,624,604
&   3,560,242\\
\hline
Sao Paolo-20\% & 14,850,165 & 10,414,500
&    4,435,665\\
\hline
Sao Paolo-25\% & 15,322,125  & 10,067,179
&  5,254,946\\
\hline
Sao Paolo-30\% & 15,941,078  & 9,832,657
&  6,108,421\\
\hline
Teruel-5\% & 4,579,237 & 3,602,391
&  976,846\\
\hline
Teruel-10\% & 5,218,344  & 3,591,633
&  1,626,711\\
\hline
Teruel-15\% & 5,881,409  & 3,526,322
&  2,355,087\\
\hline
Teruel-20\% & 6,330,491  & 3,489,388
&   2,841,103\\
\hline
Teruel-25\% & 7,098,750  & 3,604,679
& 3,494,071\\
\hline
Teruel-30\% & 8,159,407  & 4,731,502
&  3,427,905\\
\hline
Valencia-5\% & 3,757,466  & 2,995,307
&  762,159\\
\hline
Valencia-10\% & 4,190,315 & 3,001,785
& 1,188,530\\
\hline
Valencia-15\% & 4,765,503  & 2,961,888
&  1,803,615\\
\hline
Valencia-20\% & 5,057,464  & 2,928,309
&  2,129,155\\
\hline
Valencia-25\% & 5,665,209  & 2,952,889
&   2,712,320\\
\hline
Valencia-30\% & 6,535,266  & 3,667,503
&  2,867,763\\
\hline

\end{tabular}
\end{table}

Following all the aforementioned steps, we generated 24 datasets, each described in Table \ref{tab:table3}. This table provides details on the number of samples, along with the counts of both normal and malicious samples in each dataset. After completing the simulations, the next stage involves data preparation for training and testing the proposed system. For each node in the simulation, we aggregate the count of received packets from various nodes in 1-second intervals. This data serves as the foundation for feature engineering, focusing on the last 10 seconds of packet reception history, which results in 10 distinct features. These features encapsulate the packet counts received from different senders over time, supplying valuable information for training and evaluating tree-based models.

In essence, if there is an anomalous change in the number of packets received from each node per second, this alteration is highly likely to be detected within the initial second of observation. Our results substantiate this notion, underscoring the effectiveness of early detection in the first second of observation. This method introduces a novel approach to data preprocessing and feature engineering, aiming to enhance detection rates, achieve dimensionality reduction, and enhance accuracy while minimizing false alarms and time complexity.

In contrast to traditional feature selection methods, which can introduce additional overhead and complexity, our approach involves simplified and expedited calculations. These calculations allow us to directly provide results to the models, eliminating the need for extensive feature selection procedures. Consequently, this approach offers a more efficient and streamlined process for data preparation, enhancing model training without compromising performance.

Given these considerations, a dataset is created for each vehicle, where each dataset consists of \textit{n} rows, corresponding to the aforementioned time periods.
\begin{itemize}

\item   Let \textit{line($t_i$)} be the row created for time period \textit{$t_i$}. 
\item 	Let \textit{a} be a randomly chosen integer, \textit{a}  represents the number of features for the dataset. We set \textit{a = 10} in our work.
\item  Then 
$\text{Line}(t_i)$ contains $\textit{Count}(t_i), \textit{Count}(t_{(i-1)}), \\ \textit{Count}(t_{(i-2)}), \ldots, \textit{Count}(t_{(i-(a-1))})$ as the features.
\item Each line is labelled as a positive exemplar if the attack is in progress at time period \textit{$t_i$}, negative otherwise.
\end{itemize}

Since we intend to perform both training and testing for each vehicle separately, we need to segment the datasets based on each vehicle's unique ID. This segmentation allows us to run the method and obtain results for each vehicle individually. In Table \ref{tab:time}, we present the average time taken by each node for data preprocessing and dataset preparation for training and testing. 
\begin{table}[!t]
\caption{Time (in Seconds) }\label{tab:time}
\centering
\begin{tabular}{|c||c||c||c|}
\hline
Dataset & Prepro.
& Attack Onset De.
& Malicious Node De. \\
\hline
Manhattan-5\% & 0.70 &    5.46
&     0.004\\
\hline
Manhattan-10\% & 0.70 &   6.2 & 0.004\\
\hline
Manhattan-15\% & 0.70 &  6.26&   0.004\\
\hline
Manhattan-20\% & 0.71	&5.85	&0.005\\
\hline
Manhattan-25\% & 0.71&	5.78	&0.005\\
\hline
Manhattan-30\% & 0.72&	5.86	&0.005\\
\hline
Sao Paolo-5\% &0.72	&6.26&	0.008\\
\hline
Sao Paolo-10\% & 0.72	&5.96	&0.008\\
\hline
Sao Paolo-15\% & 0.71&	5.88&	0.008\\
\hline
Sao Paolo-20\% & 0.72&	5.59&	0.008\\
\hline
Sao Paolo-25\% & 0.73&	5.41&	0.008\\
\hline
Sao Paolo-30\% & 0.74	&5.65&	0.009\\
\hline
Teruel-5\% & 0.70&	5.93	&0.004\\
\hline
Teruel-10\% & 0.70&	6.62&	0.005\\
\hline
Teruel-15\% & 0.70&	6.94&	0.005\\
\hline
Teruel-20\% & 0.70&	5.85&	0.005\\
\hline
Teruel-25\% & 0.71&	5.79	&0.005\\
\hline
Teruel-30\% & 0.71&	5.86&	0.006\\
\hline
Valencia-5\% & 0.69&	6.57&	0.004\\
\hline
Valencia-10\% & 0.69	&6.51	&0.004\\
\hline
Valencia-15\% & 0.70&	5.73&	0.004\\
\hline
Valencia-20\% & 0.70&	5.69&	0.004\\
\hline
Valencia-25\% & 0.70	&5.86	&0.004\\
\hline
Valencia-30\% & 0.71	&6.46&	0.005\\
\hline
\end{tabular}
\end{table}

Determining what constitutes \textit{"enough"} data for training is crucial for ML algorithms. To this end, we employed the Synthetic Minority Over-sampling Technique (SMOTE) to address this challenge by augmenting the dataset. The goal is to ensure that the learning model is provided sufficient and representative data, optimizing its training for improved performance. This approach aligns with our overarching research objective of achieving robust and effective model outcomes across various VANET scenarios.

\subsection{The Proposed System: ADVENT}  
In our evaluation of our proposed system, ADVENT, the process starts with each vehicle (node) training an XGBoost model using its local data. Subsequently, the node sends the resulting trees, accompanied by a unique Client ID (CID), to the central server. The server then consolidates these trees and, in the following steps, updates the global model weights (W) to improve efficiency while safeguarding the privacy of individual vehicle data. In the event of any ongoing attacks during these rounds, ADVENT triggers the Malicious Node Detection component. To identify potential threats, each vehicle calculates a threshold based on the count of inbound connections from neighboring nodes, labeling those vehicles exceeding this threshold as \textit{Suspected}. The lists of \textit{Suspected} vehicles are then collectively shared with the server to create an aggregated list, and the Malicious Node list is subsequently distributed to all nodes for proactive attack prevention and protection against receiving compromised data.

\subsubsection{Data Preprocessing Component}
\label{pre}

Each node in the network receives packets from all neighbors, and two different data presentations are required for the two main following steps in detecting malicious behavior.

For attack onset detection, each node actively computes the number of received packets from all nodes in the neighborhood within a predefined time interval, creating a history of network behavior. This history is then used to detect the onset of an attack. Each node actively and continously prepares this history and utilizes it for attack onset detection.

The second data type needed is the number of received packets from each node and the Mean Absolute Deviation (MAD) calculated based on these numbers. Each node computes MAD based on the received packets from each node in intervals of 1 minute until it detects the onset of an attack. Once an attack onset is detected, nodes run the procedure in 10-second intervals to identify and block malicious nodes. Afterward, they revert to 1-minute intervals until another attack onset is detected. The intervals can be adjusted based on network conditions to achieve optimal efficiency.

\subsubsection{Attack Onset Detection Component}
\label{A_ONSET}
\begin{figure}[!t]
\centering
\includegraphics[width=4.7in]{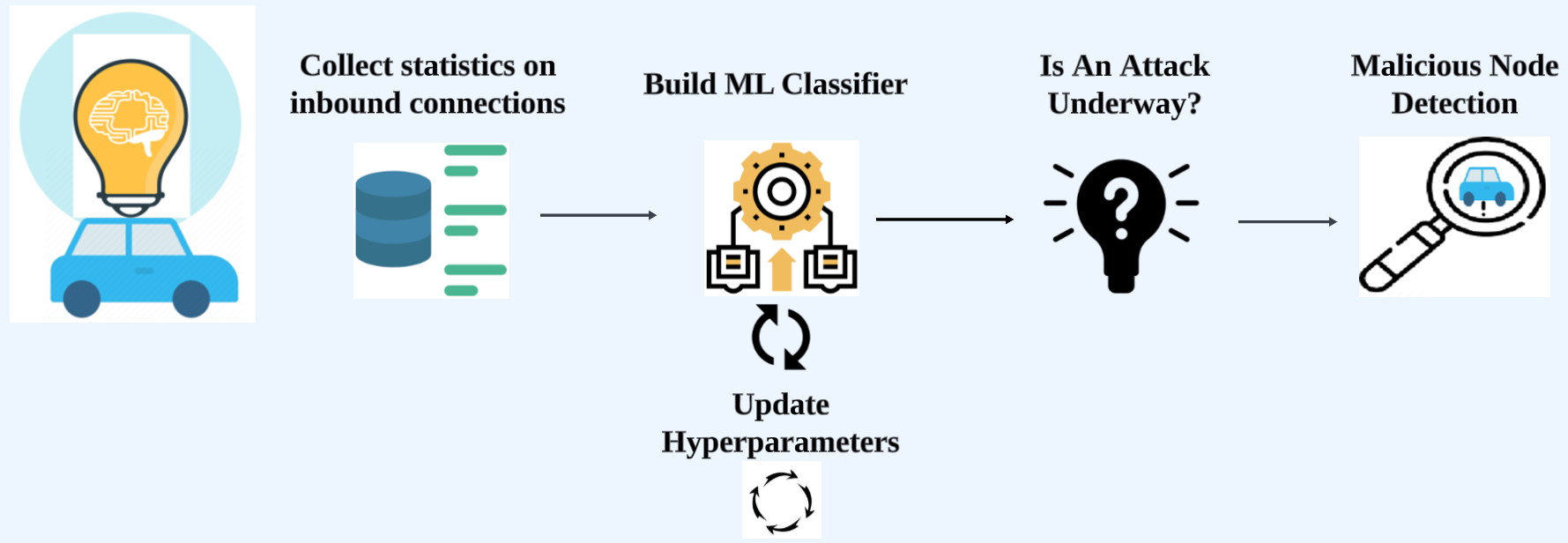}
\caption{The Attack Onset Detection system}
\label{A_ONSET_FG}
\end{figure}

In Figure \ref{A_ONSET_FG}, the process of attack onset detection commences with the acquisition of the preprocessed data, as elaborated in \ref{pre}, where the number of received inbound connections are computer over 1-second windows over the last 10 seconds. Subsequently, the XGBoost classifier is built using these data. 

At time t=0 (time 0 seconds), all nodes train their local XGBoost tree ensemble and send the trees to the server. The server aggregates the tree ensemble and initializes a Convolutional Neural Network (CNN). After receiving the aggregated tree ensemble, nodes calculate prediction outcomes given the aggregated trees on their local data samples, and these outcomes become inputs for the CNN.

Notably, clients only build XGBoost models at round 0, and the aggregated tree ensemble remains fixed after round 0. For federated training of the one-layer 1D CNN after round 1, the protocol follows a standard FL algorithm, updating weights for 9 rounds, finalizing the model each node can use during its travel. Results of our simulations indicate that it takes around 6 seconds before nodes can start detecting the onset of an attack.

\subsubsection{Malicious Node Detection Component}
\label{M_DE}
\begin{figure}[!t]
\centering
\includegraphics[width=4.7in]{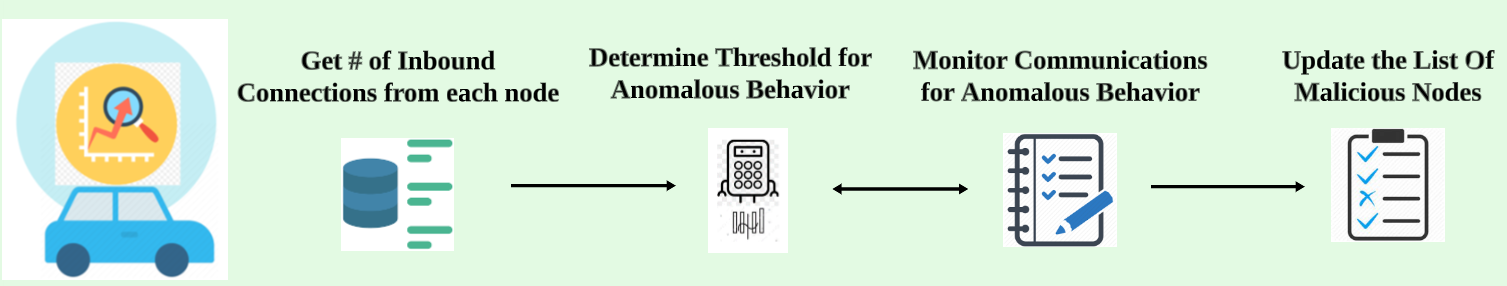}
\caption{The Malicious Node Detection system}
\label{M_DE_FIG}
\end{figure}

Figure \ref{M_DE_FIG} presents the overview of the Malicious Node Detection component, which becomes active when the Attack Onset Detection component detects an initiation of an attack. This component's first step is to get the number of inbound connections from each node. Subsequently, it moves on to establishing the threshold for identifying anomalous behavior, a process performed based on the Median Absolute Deviation (MAD) algorithm. \cite{ours,ours2,MDASTI}

MAD employs an anomaly detection technique, specifically leveraging Median Absolute Deviation as a statistical model for detecting outliers or anomalies. MAD computes the median of the data and considers the absolute differences between each data point and the median. Through MAD, the system scrutinizes the number of connection requests received from neighboring vehicles, seeking indications of anomalies like flooding attacks or DDoS attacks perpetrated by malicious vehicles. 

Advent calculates MAD using Equation \eqref{eqn:mad}. It should be noted here that the median of each series is calculated and multiplied by 'b = 1.4826' which ensures the estimator's consistency with respect to the parameter of interest \cite{leys2013detecting}.

\begin{equation}
\label{eqn:mad}
MAD = b M_i(|x_i - M_j (x_j)|)
\end{equation}

By utilizing MAD, ADVENT identifies anomalies in the number of requests, abnormal behaviors in vehicular networks, contributing to the detection of DDoS attacks.

The next critical step involves establishing a rejection threshold, denoted as \textit{tr}. Values exceeding this threshold are classified as outliers or anomalies.
To determine 'tr' and evaluate different exclusion criteria (ce),based on prior work \cite{ours} the value of \textit{ce = 3.0} is set. Equation \eqref{eqn:tr} is used for calculating the rejection threshold:

\begin{equation}
\label{eqn:tr}
M - ce * MAD< t_r< M + ce * MAD
\end{equation}

where 
\textit{tr} represents the rejection threshold, either negative or positive, used to identify outliers.
\textit{MAD} represents the calculated Median Absolute Deviation value.
\textit{ce} represents the exclusion criterion, and 
\textit{M} represents the median value of the original classified series.

Once the rejection threshold \textit{tr} is determined based on the chosen exclusion criterion, it is applied to the anomaly detection process. Data points falling outside of this threshold are classified as outliers or anomalies \cite{MDASTI}. Determining the threshold enables the detection mechanism to distinguish between normal and abnormal behavior based on data dispersion. This facilitates the identification of potential malicious nodes or unusual patterns within the VANET environment.

Up to this point, we have introduced a method for detecting malicious nodes. However, there is a crucial concern, as discussed by Baharlouei et al. in \cite{ours2}. Using this method, the results are heavily influenced by the chosen grid pattern and each node's experience of encountering attackers. Specifically, if a vehicle is located far from an attacker or if there are obstructions that hinder packet reception, that vehicle may not receive sufficient packets from the attacker to identify it as malicious. Consequently, it might not include the attacker in its suspicious list. This could result in a high FNR and affect the overall F1-score as well as DR.

To address this challenge, ADVENT utilizes Federated Learning, which leverages each node's unique experience and allow vehicles to share these experiences with one another. This collaborative approach ensures that all vehicles possess the most accurate and comprehensive blacklist of potential malicious nodes, enabling them to interact with their neighbors more effectively. 

\subsubsection{Federated Learning Component}
Federated Learning (FL) is an innovative approach to collaborative machine learning, gaining recognition for its role in preserving data privacy and optimizing data-driven applications, in particular within VANETs\cite{lu2020federated}. FL differs from traditional centralized approaches by enabling model training on local devices, such as vehicles in VANETs, allowing them to learn from their data while aiming to ensure data privacy\cite{alam2022federated}. In FL, only the model updates are shared among the nodes, eliminating the need to exchange raw data to preserve data privacy. This decentralized approach offers multiple advantages for VANETs, including enhanced security, reduced communication overhead, and increased adaptability\cite{electronics12102287}. Furthermore, FL promotes the development of robust and accurate models by aggregating diverse local knowledge, empowering vehicles to collaboratively identify malicious nodes, predict traffic patterns, and enhance the overall performance of a vehicular network \cite{electronics12102287}

Indeed, FL can play a crucial role in improving the FNR within the Malicious Node Detection system of ADVENT. By employing FL to aggregate lists and harness the collective experience of various nodes, a comprehensive overview of the network and the behavior of all vehicles can be aggregated. This collaborative approach allows us to detect malicious nodes, reduce the likelihood of false negatives, and improve the accuracy of the detection process.

\subsubsection*{\bf Federated Learning in Attack Onset Detection}
In this step instead of transmitting raw data from each vehicle for processing, we opt for sharing models and trees during the training process, aligning with the principles of FL. Our research leverages FedXGBllr \cite{FLXGBOOST}, which deviates from existing methods of federated XGBoost. Rather than relying on gradient sharing, the FedXGBllr approach takes a novel route by minimizing per-node level communication frequency and mitigating privacy concerns. This system goes beyond privacy enhancement and also optimizes communication efficiency. It achieves this by enabling the learning rates of the aggregated tree ensembles to be adjusted, ensuring the security of sensitive information during model training optimization. \cite{FLXGBOOST}.

\begin{figure}[!t]
\centering
\includegraphics[width=4.7in]{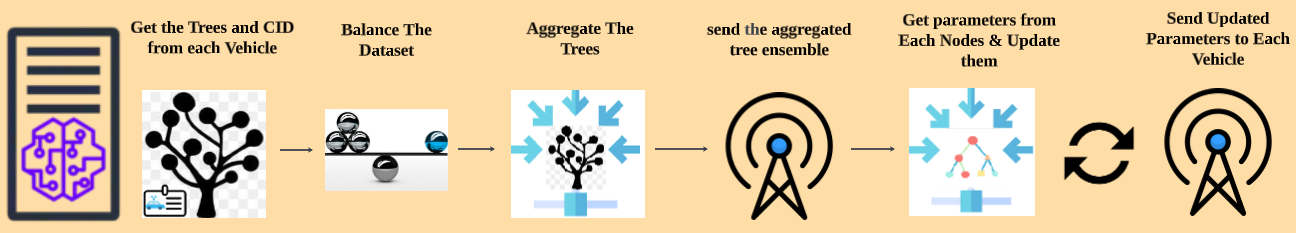}
\caption{The Federated Learning-Attack Onset Detection system}
\label{FL_A_ONSET}
\end{figure}

Figure \ref{FL_A_ONSET} provides a visual representation of this component of the system, illustrating how the server receives the trees and Client IDs (CIDs) from the vehicles. The process commences with the server's aggregation and broadcast of the trees and associated parameters. This methodology ensures the creation of a reliable and robust model without imposing excessive processing demands on the vehicles or the need to share extensive data among vehicles. Next, we will delve into a detailed discussion of the Federated Learning method we employed.

\subsubsection*{\bf In the Initial Round \textit{(t = 0)}}
Upon receiving the trained XGBoost tree ensembles and Client IDs (CIDs) from vehicles that have trained their local XGBoost trees using available data, the server undertakes the aggregation of these tree ensembles. This critical step is instrumental in facilitating the training of a global federated XGBoost model. The server proceeds by:

\begin{itemize}
\item{Sorting and broadcasting the aggregated XGBoost tree ensemble along with corresponding CIDs to all clients.}
\item{Initializing \textit{w0} for the one-layer 1D CNN.}
\item{Each client receives the broadcasted aggregated XGBoost tree ensemble and proceeds to evaluate these ensembles with their localized dataset. This evaluation yields prediction outcomes generated by all trees for each data sample.}
\end{itemize}
\subsubsection*{\bf For Subsequent Rounds(R) \textit{(t = 1 to R)}}
\begin{itemize}
\item{The server designates the clients who will participate.}
\item{Global model parameters, represented as \textit{wt}, are broadcast to all clients by the server.}
\item{In parallel fashion, each client updates its local model parameters to \textit{wk(t+1)} by performing a Client Update operation, utilizing the received parameters.}
\item{Subsequently, the server aggregates the updated models from all clients to calculate the new global model parameters, \textit{wk(t+1)}.}
\end{itemize}

This iterative process unfolds over subsequent rounds, ensuring that the global model continually refines itself with each step. This sophisticated and collaborative approach aligns seamlessly with the principles of Federated Learning, promoting effective global model improvement while keeping individual data private. It also offers the advantage of faster training with local data, removing the need for servers in the training process \cite{MTH}. 

\subsubsection*{\bf Federated Learning in Malicious Node Detection}
\begin{figure}[!t]
\centering
\includegraphics[width=4.7in]{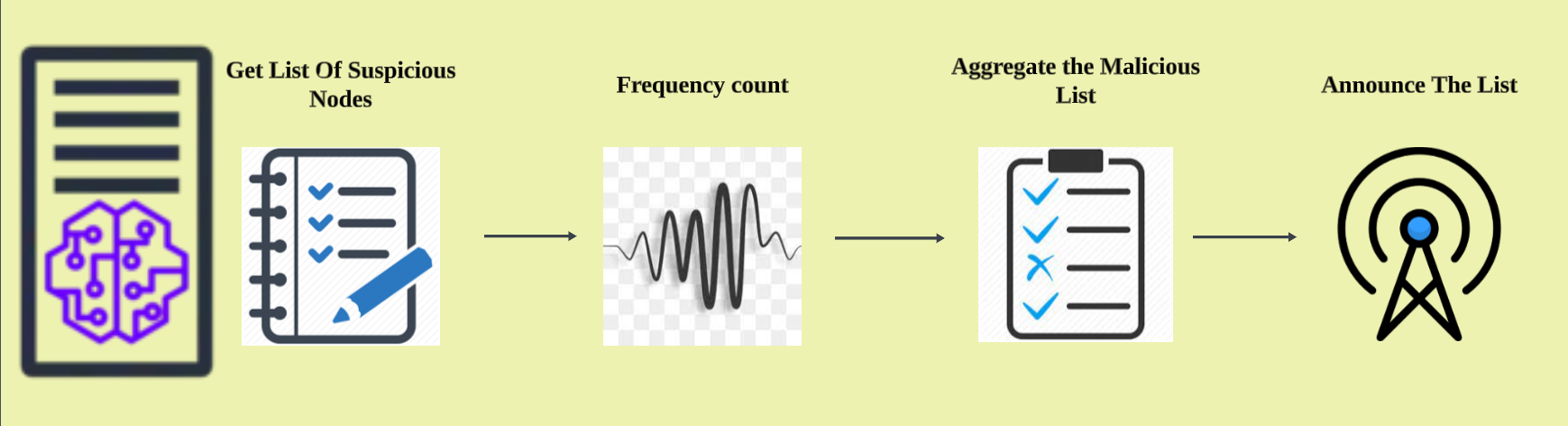}
\caption{The Federated Learning-Malicious Node Detection system}
\label{FL_M_fig}
\end{figure}

Fig. \ref{FL_M_fig} illustrates the application of the Federated Learning approach to the detection of malicious nodes. In previous work by Baharlouei et al. as described in \cite{ours2}, the MDASTI method was utilized for malicious node detection, and it highlighted a significant issue concerning the False Negative Rate (FNR). The variation in each vehicle's experiences within a network depends largely on factors such as the number of attackers and the grid pattern, as well as geographical characteristics of the simulations. Since these factors are not controllable and can introduce considerable variability into the results, it becomes imperative to take measures to mitigate these side effects. To address this, we propose a novel method that aims to combine the collective experience of all nodes, allowing vehicles to view the network from the perspective of other vehicles. After the Malicious Node Detection component has been executed in a vehicle and list of suspicious actors is generated, this new component is called into action. In this component, as depicted in Fig. \ref{FL_M_fig}, the server receives the suspicious lists from all vehicles and proceeds with the following steps:

\begin{itemize}
\item{Count the frequency of each reported vehicle \textit{i}, denoted as \textit{Fi}}
\item{If the frequency of a reported vehicle, Fi, is higher than the predefined threshold "TH" set for the network, then include that vehicle in the list of malicious nodes}
\item{Transmit the malicious node list to all vehicles in the network}
\end{itemize}

This method aids nodes in identifying malicious nodes within the network even before they receive a substantial number of inbound connections or come under attack from malicious nodes. It significantly contributes to the reduction of the False Negative Rate (FNR), allowing nodes to base their decisions on the maximum connection requests sent by each node in the network rather than the proportion they receive. The inclusion of this component in the system has led to several positive achievements: Firstly, a notable reduction in FNR, coupled with an increase in DR and F1-score, as previously discussed in our prior work. \cite{ours2}; secondly, nodes are now capable of predicting attacks before the influx of inbound connections, effectively avoiding interactions with malicious nodes; thirdly, it has minimized the impact of differing geographical patterns and varying numbers of attackers on the network.
\label{FL}
\subsection{Performance Metrics}
In our study, we employed the standard evaluation metrics, namely Detection Rate (DR), False Alarm Rate (FAR), False Negative Rate (FNR), and F1-score, which are commonly used to assess the performance of a machine learning model. These metrics were used to evaluate the systems accuracy in the context of attack onset detection and malicious node identification. These metrics are defined as follows: Detection Rate (DR) measures the proportion of true positive instances correctly identified, False Alarm Rate (FAR) quantifies the proportion of false alarms, False Negative Rate (FNR) evaluates the rate of false negatives, and the F1-score is a balanced measure of precision (positive predictive value) and recall (Sensitivity or True Positive Rate). To calculate these metrics, we used the following formulas, where True Positive (TP) represents the number of correctly identified positive instances, True Negative (TN) denotes the number of correctly identified negative instances, False Negative (FN) indicates the number of instances mistakenly classified as negative, and False Positive (FP) signifies the number of instances mistakenly classified as positive. These metrics are instrumental in evaluating the effectiveness of our proposed system.
\begin{equation}
\label{eqn:DR}
DR = TP / (TP + FN)
\end{equation}

\begin{equation}
\label{eqn:FAR}
FAR = FP / (FP + TN)
\end{equation}

\begin{equation}
\label{eqn:FNR}
FNR = FN / (TP + FN)\\
\end{equation}

\begin{equation}
\label{eqn:Recall}
Recall = TP / (TP + FN)\\
\end{equation}

\begin{equation}
\label{eqn:Prec}
Precision = TP / (TP + FP)\\
\end{equation}

\begin{equation}
\label{eqn:F1}
\textit{F1-score} = 2 * (Precision * Recall) / (Precision + Recall)
\end{equation}

\subsection{Results}
In ADVENT, once each vehicle has trained its local model and shared the trees with the server, the server aggregates these trees and distributes them to all vehicles during round 0. Subsequently, only the model weights are updated, following the methodology outlined in \cite{FLXGBOOST}. To maintain consistency with this reference in our simulations, we have set the number of rounds to 10 and Epochs to 100, with a batch size of 64. The XGBoost parameters have been configured based on the default values specified in \cite{FLXGBOOST}. After training each vehicle's model and optimizing it, we conducted testing using the unseen data for each vehicle. The results are presented in the following figures.

\subsubsection*{\bf Results of Attack Onset Detection}
\begin{figure}[!t]
\centering
\includegraphics[width=4.7in]{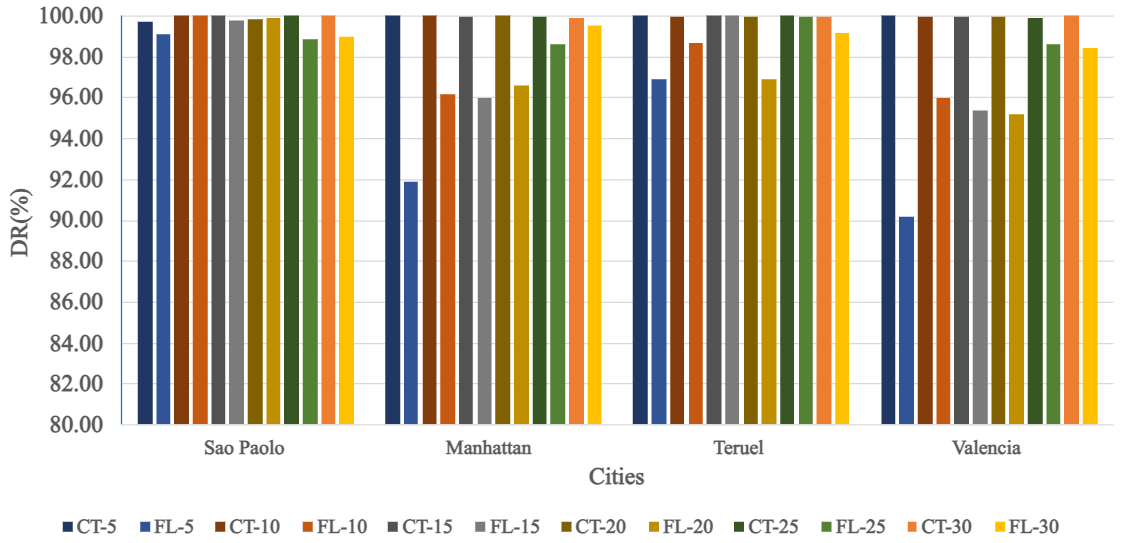}
\caption{Attack Onset Detection-Detection Rate}
\label{AD_DR_fig}
\end{figure}

: Attack Onset detection harnesses XGBoost to identify the initiation of attacks.
We began our analysis by considering the DR, as depicted in Figure \ref{AD_DR_fig}, we compared the outcomes of employing Federated Learning as a distributed method to uphold privacy, where nodes operate and train locally based on their data. 
Contrasting these results with the centralized approach, as seen in the figure, where XGBoost is trained with 70\% of each dataset and tested with the remaining 30\%, ADVENT achieved more realistic outcomes. This approach considers the variation in results across datasets, reflecting the predictability of obtaining diverse outcomes in datasets with varying numbers of malicious and benign samples. Over 20 datasets, it is observed the results remain consistent in most scenarios except in some cases where the performance decreases slightly. The disparity is more evident in smaller datasets where there are fewer samples to train local models for both benign and malicious instances. Sao Paolo exhibited the best results, having the largest dataset, followed by Teruel, Manhattan, and Valencia with the smallest dataset, resulting in the least favorable outcomes. Furthermore, as the number of attackers increase in each city, consequently augmenting the number of attack onset samples, we observe that results are improved, leading to a reduction in differences among datasets.

Moving on to the next metric, F1\_Score, calculated using equation \ref{eqn:F1}, we observe a pattern similar to the DR results. Sao Paolo consistently demonstrates the best performance due to its larger dataset, allowing for ample samples to train local models in each node. Additionally, across all nodes, a higher number of attackers results in a higher F1\_Score, indicating improved performance as more samples are available for training. However, it's noteworthy that with 55 attackers, the F1\_Score does not yield the optimal results.

\begin{figure}[!t]
\centering
\includegraphics[width=4.7in]{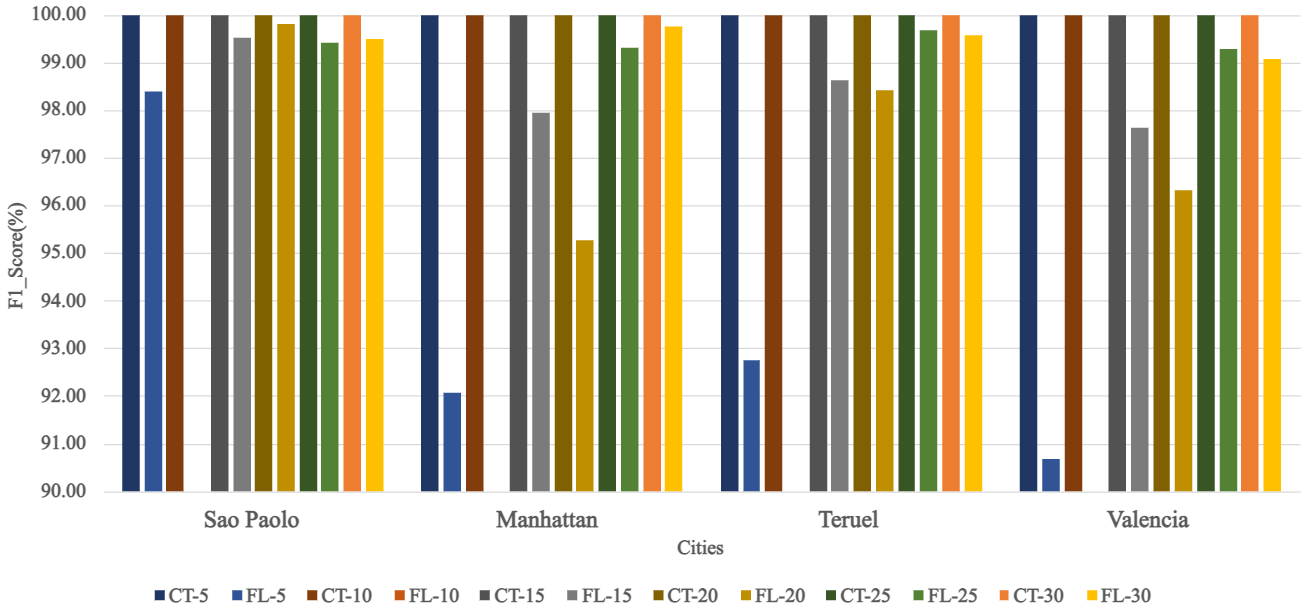}
\caption{Attack Onset Detection-F1-Score}
\label{AD_F1_fig}
\end{figure}

\begin{figure}[!t]
\centering
\includegraphics[width=4.7in]{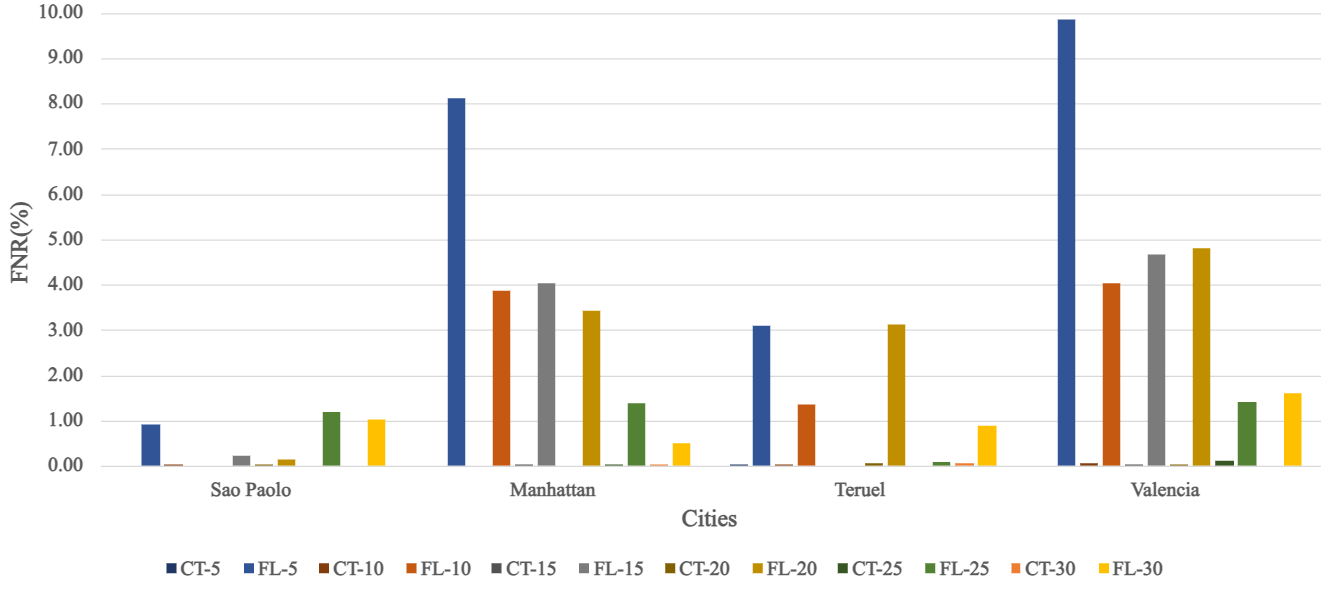}
\caption{Attack Onset Detection-False Negative Rate}
\label{AD_FNR_fig}
\end{figure}

In Figure \ref{AD_FNR_fig}, the discernible surge in the FNR from an average of 0.02\% to 2.48\% can be directly attributed to a lower number of samples available for training the model. This effect is particularly pronounced in the Manhattan and Valencia scenarios, where the datasets are smaller, especially when dealing with 5\% of attackers. The reduced sample size in these instances contributes to the deterioration of FNR results, highlighting the critical role of dataset size in influencing model performance especially when we try to train the model locally and provide privacy. 

\begin{figure}[!t]
\centering
\includegraphics[width=4.7in]{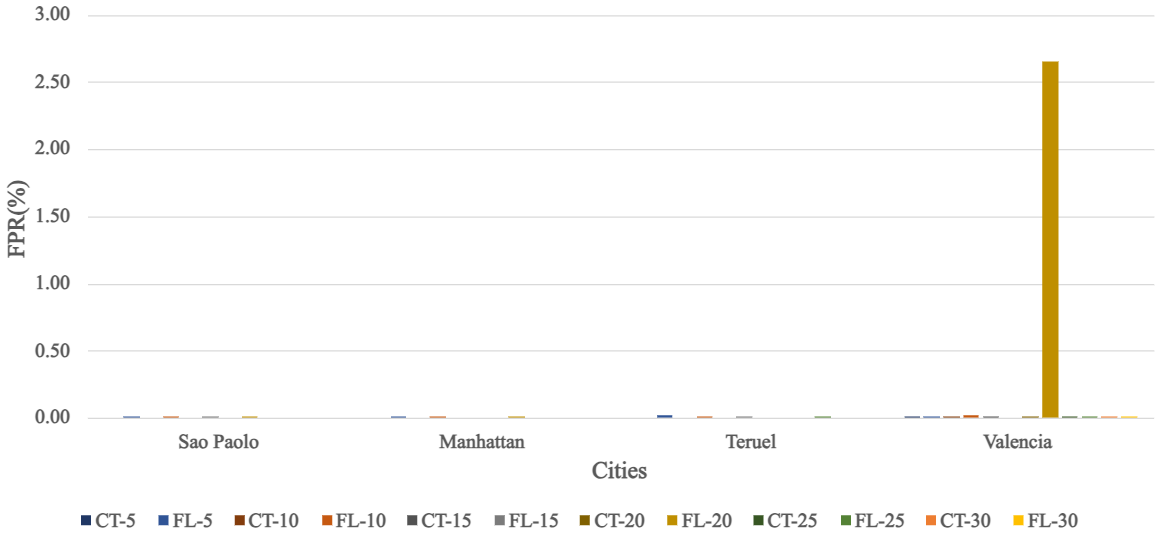}
\caption{Attack Onset Detection-False Positive Rate}
\label{AD_FPR_fig}
\end{figure}

In Figure \ref{AD_FPR_fig}, a marginal shift in the FPR is evident, moving from 0\% to 0.11\%. This subtle change underscores the reliability of the model in accurately detecting the attack, emphasizing its trustworthiness even under these comparative conditions.

The results seen so far seem to suggest that the number of samples per node significantly influences both the training process and the quality of the produced trees. We, thus, delved deeper to explore the relationship between dataset size and detection ability. Table \ref{tab:N/M samples} outlines the number of malicious and normal samples in both the train and test datasets after dataset preparation and preprocessing. The last column provides the Normal/Malicious ratio, revealing that Sao Paolo attains the best results with the lowest ratio. Further analysis in Table \ref{tab:table3} indicates that Sao Paolo also possesses the largest datasets. Additionally, the ratio varies with changes in the number of attackers, being lower in scenarios with a higher number of attackers.

\begin{table}[htbp]
  \centering
  \caption{\# of Malicious and Normal samples in Datasets after preprocessing the datasets}
    \begin{tabular}{lrrrrr}
    \multirow{2}[0]{*}{DataSet} & \multicolumn{2}{l}{Malicious} & \multicolumn{2}{l}{Normal} & \multicolumn{1}{l}{Train} \\
          & \multicolumn{1}{l}{Test} & \multicolumn{1}{l}{Train} & \multicolumn{1}{l}{Test} & \multicolumn{1}{l}{Train} & \multicolumn{1}{l}{Normal/Mal} \\
    Manhattan-5\%  & 518   & 1418  & 394042 & 914658 & 645.03 \\
    Manhattan-10\% & 700   & 1586  & 393860 & 914254 & 576.45 \\
    Manhattan-15\%  & 745   & 1682  & 393815 & 914158 & 543.49 \\
    Manhattan-20\%  & 762   & 1719  & 393798 & 914121 & 531.77 \\
    Manhattan-25\%  & 785   & 1774  & 393775 & 914066 & 515.26 \\
    Manhattan-30\%  & 810   & 1830  & 393750 & 914010 & 499.46 \\
    Sao Paolo-5\% & 1331  & 3035  & 384589 & 892645 & 294.12 \\
    Sao Paolo-10\%  & 1319  & 2980  & 384601 & 892700 & 299.56 \\
    Sao Paolo-15\%  & 1327  & 2997  & 384593 & 892683 & 297.86 \\
    Sao Paolo-20\%  & 1345  & 3035  & 384575 & 892645 & 294.12 \\
    Sao Paolo-25\%  & 1354  & 3053  & 384566 & 892627 & 292.38 \\
    Sao Paolo-30\%  & 1364  & 1364  & 384556 & 384556 & 281.93 \\
    Teruel-5\%  & 646   & 1456  & 390674 & 906824 & 622.82 \\
    Teruel-10\% & 738   & 1670  & 390582 & 906610 & 542.88 \\
    Teruel-15\%  & 783   & 1776  & 390537 & 906504 & 510.42 \\
    Teruel-20\%  & 807   & 1818  & 390513 & 906462 & 498.60 \\
    Teruel-25\%  & 849   & 1913  & 390471 & 906367 & 473.79 \\
    Teruel-30\% & 905   & 2038  & 390415 & 906242 & 444.67 \\
    Valencia-5\%  & 426   & 1305  & 389814 & 904793 & 693.33 \\
    Valencia-10\%  & 672   & 1536  & 389568 & 904224 & 588.69 \\
    Valencia-15\%  & 709   & 1607  & 389531 & 904153 & 562.63 \\
    Valencia-20\%  & 728   & 1647  & 389512 & 904113 & 548.95 \\
    Valencia-25\%  & 769   & 1738  & 389471 & 904022 & 520.15 \\
    Valencia-30\% & 818   & 1849  & 389422 & 903911 & 488.86 \\
    \end{tabular}%
  \label{tab:N/M samples}%
\end{table}%

\begin{figure}[!t]
\centering
\includegraphics[width=4.7in]{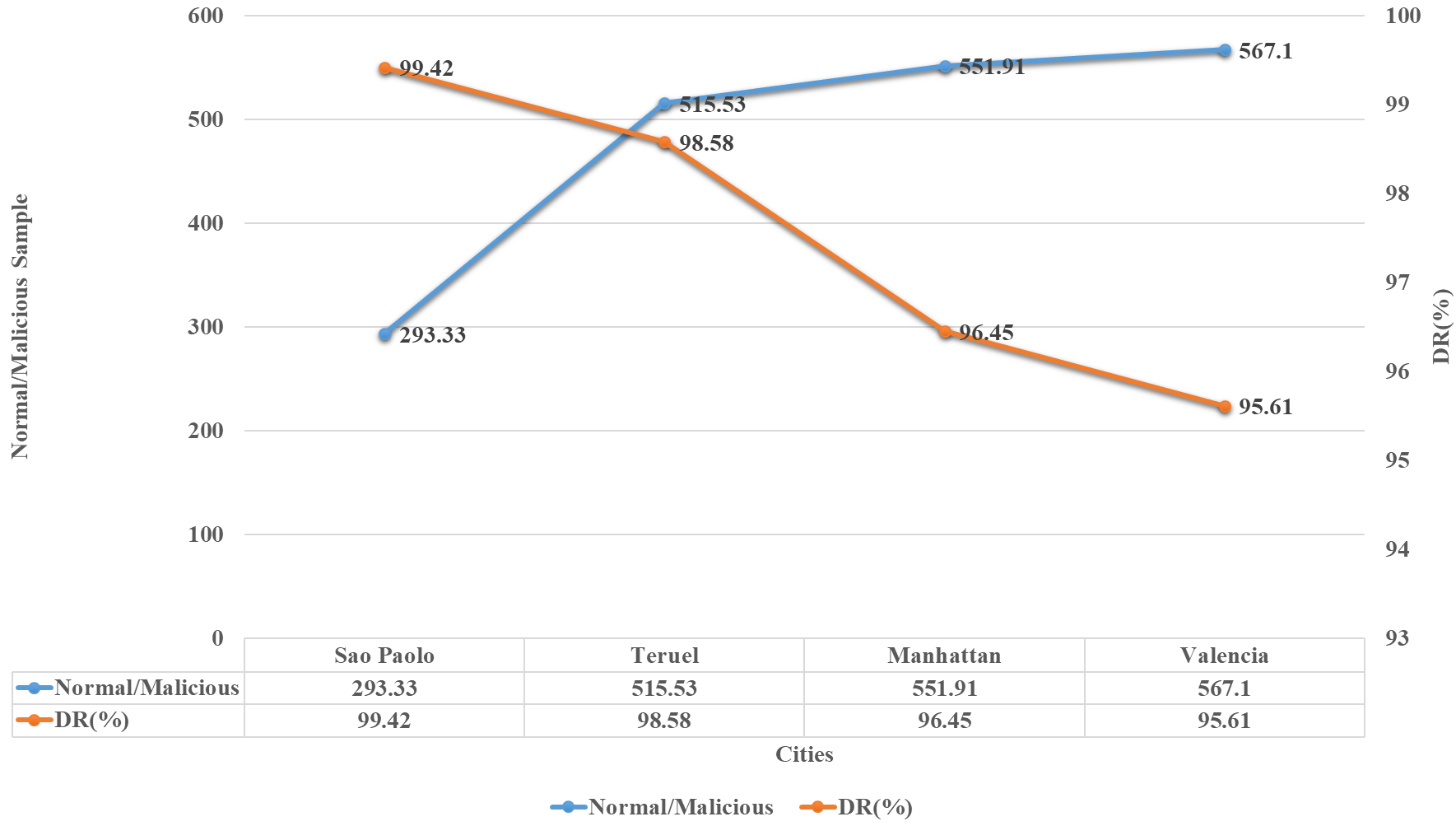}
\caption{Attack Onset Detection-False Positive Rate}
\label{AD_N/M packets_fig}
\end{figure}

Further explorations are conducted in order to understand the relationship between the Normal/Malicious data ratio and the DR. As illustrated in Figure \ref{AD_N/M packets_fig},there seems to be a direct correlation between the Normal/Malicious sample ratio in each dataset and the DR. Notably, Sao Paolo, with the lowest ratio, yielded the most favorable results, while Valencia, with the highest ratio, exhibited the poorest outcomes. To mitigate this imbalance and enhance local model training, we integrate Synthetic Minority Over-sampling Technique (SMOTE) into ADVENT. SMoTE is utilized to ensure that the ratio of Normal/Malicious samples is 2.9\% across all datasets. In doing so, our aim is to evaluate the impact on the results and move closer to the performance achieved through centralized training.

Going forward, our analysis includes three approaches: Centralised Training, Federated Learning, and integrating Federated Learning with SMOTE. We present our analysis and discussions for each approach over six datasets for each city. It should be noted here that the following abbreviations are used for different scenarios in the figures hereafter: Centralized Training (CT-5), using Federated Learning (FL-5), and implementing Federated Learning with SMOTE (FLS-5) on datasets with 5\% of attackers, and so forth.

\begin{figure}
    \centering
	\subfloat[Manhattan]{\includegraphics[width=0.5\linewidth]{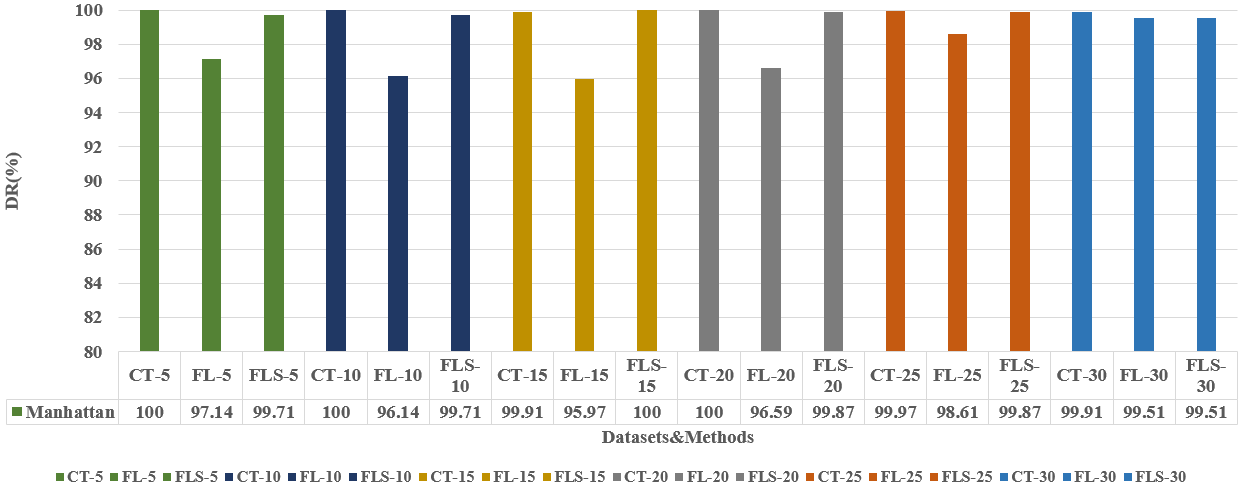}\label{ADS_M_DR_fig}}
	\subfloat[Sao Paolo]{\includegraphics[width=0.5\linewidth]{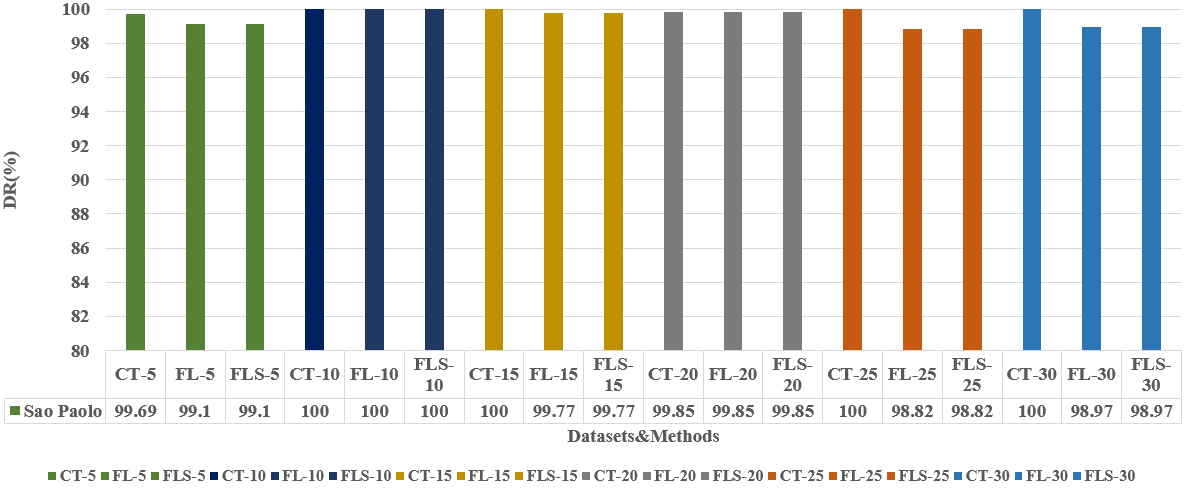}\label{ADS_S_DR_fig}} \\
	\subfloat[Teruel]{\includegraphics[width=0.5\linewidth]{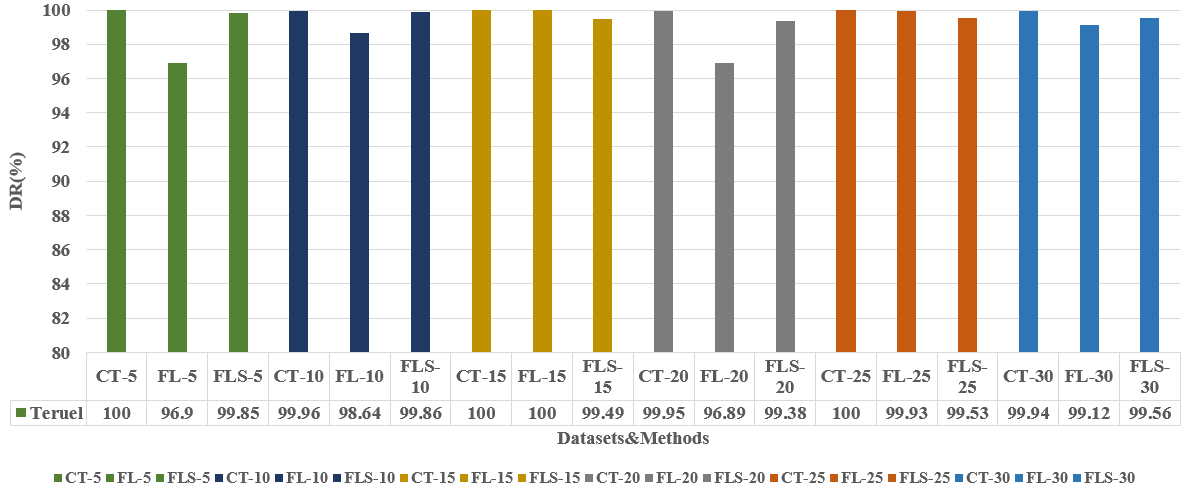}\label{ADS_T_DR_fig}}
        \subfloat[Valencia]{\includegraphics[width=0.5\linewidth]{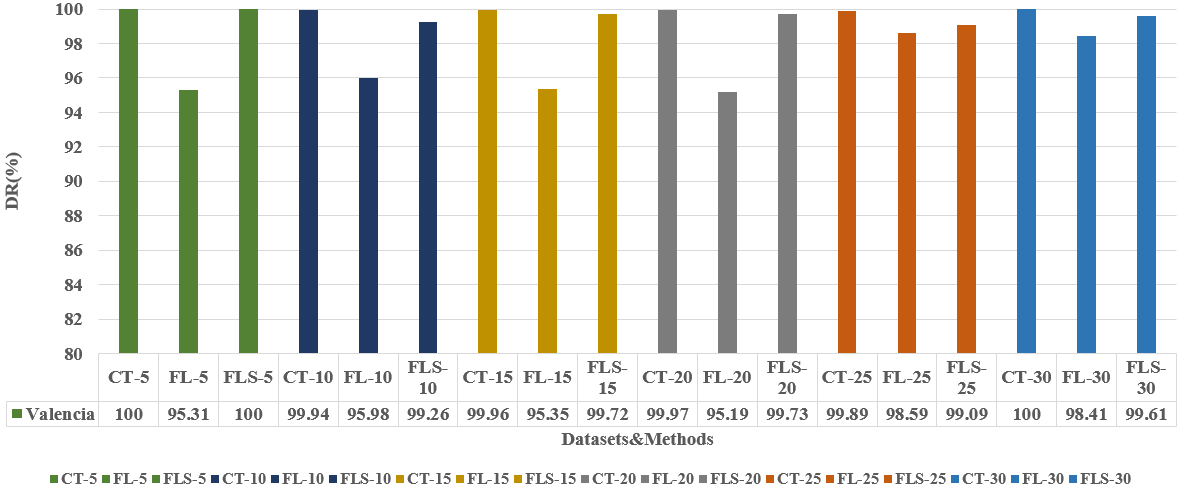}\label{ADS_V_DR_fig}}
    \caption{FLS-Attack Onset Detection-Detection Rate}
    \label{fig:FLS-DR}
\end{figure}

In Figure \ref{ADS_S_DR_fig}, the DR for Sao Paolo remained consistent as we applied the threshold based on this city, with no change in the number of training samples. However, Figure \ref{ADS_M_DR_fig} illustrates a significant impact on DR in Manhattan after applying SMOTE and increasing the number of malicious training samples. While the average DR was 99.96\% with Centralized Training and 97.33\% with the Federated Learning approach, it rose back to 99.78\% with the use of SMOTE. As shown in Figure \ref{ADS_T_DR_fig}, the DR was 99.98\% with Centralised Training, reduced to 98.58\% with Federated Learning, and then increased to 99.61\% with the application of SMOTE. Figure \ref{ADS_V_DR_fig} depicts the results for Valencia, where the DR showed the most significant difference—decreasing from 99.96\% to 96.47\% and rebounding to 99.57\% after balancing the data using SMOTE with the identified thresholds.

\begin{figure}
    \centering
	\subfloat[Manhattan]{\includegraphics[width=0.5\linewidth]{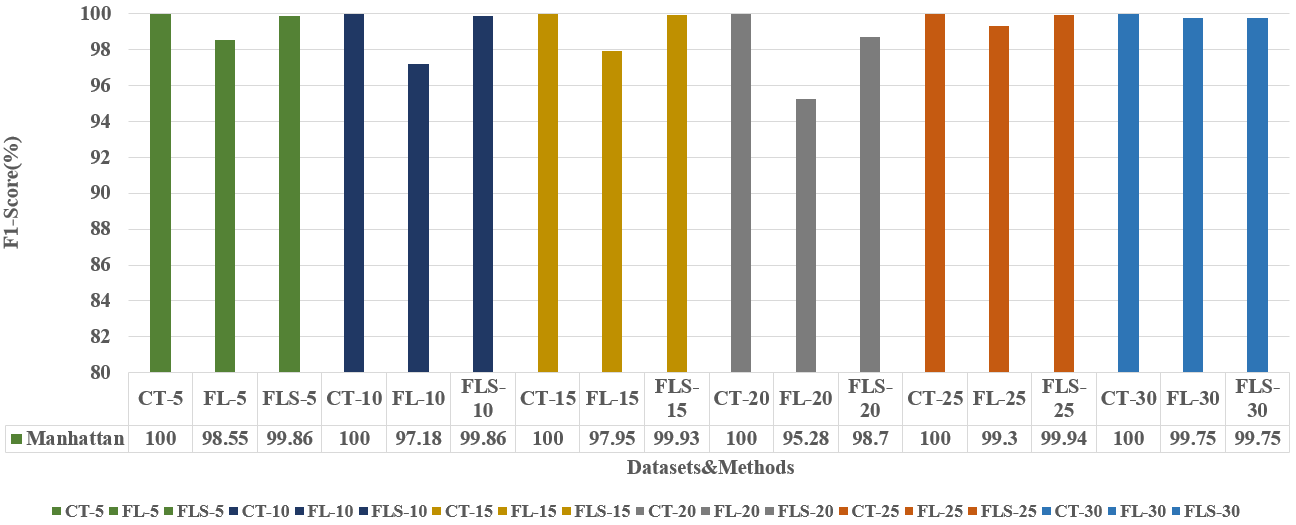}\label{ADS_M_F1_fig}}
	\subfloat[Sao Paolo]{\includegraphics[width=0.5\linewidth]{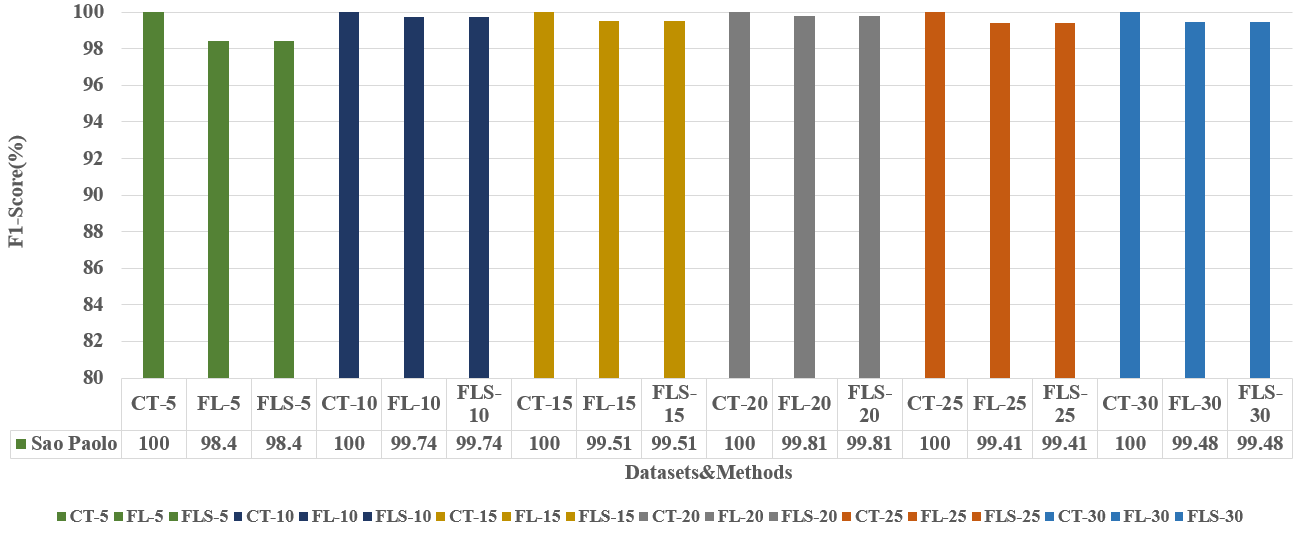}\label{ADS_S_F1_fig}} \\
	\subfloat[Teruel]{\includegraphics[width=0.5\linewidth]{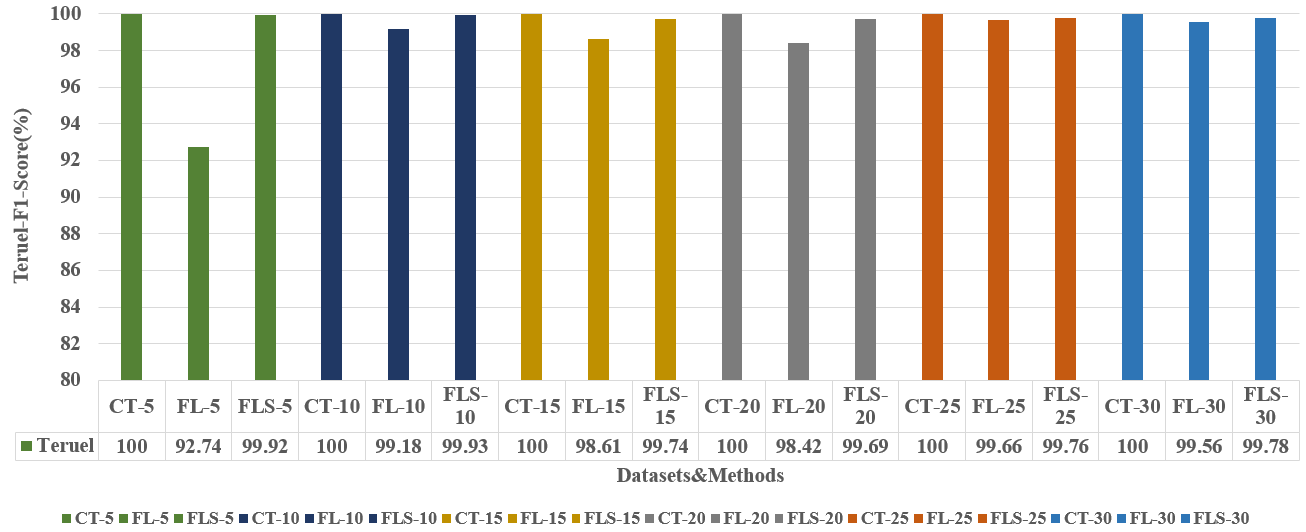}\label{ADS_T_F1_fig}}
        \subfloat[Valencia]{\includegraphics[width=0.5\linewidth]{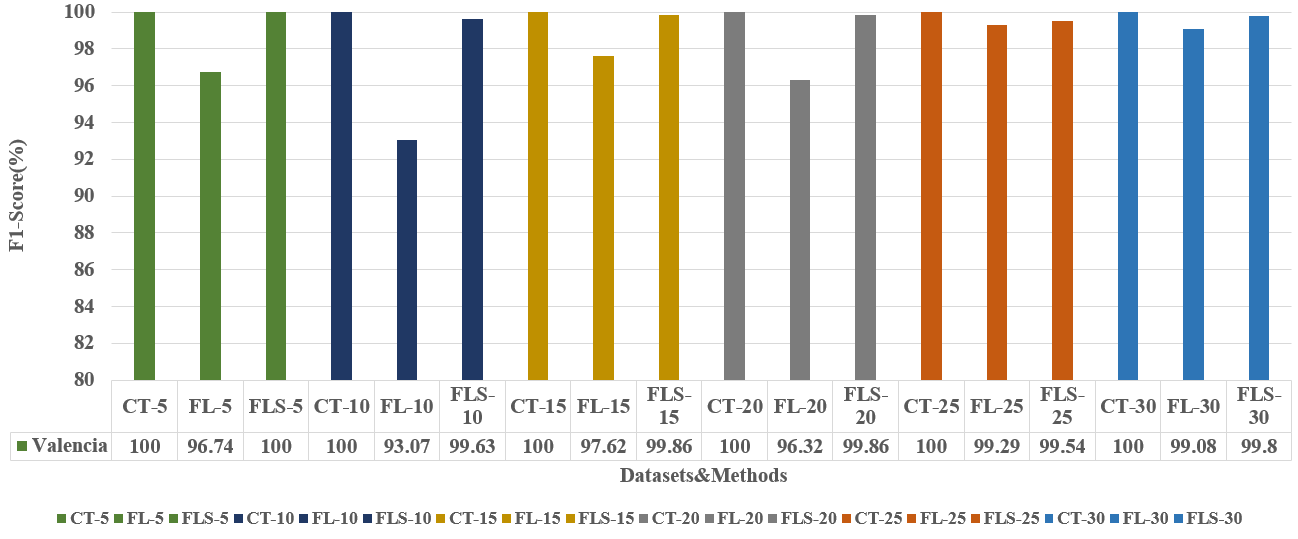}\label{ADS_V_F1_fig}}
    \caption{FLS-Attack Onset Detection-F1-Score}
    \label{fig:FLS-F1}
\end{figure}

In Figure \ref{ADS_M_F1_fig} through \ref{ADS_V_F1_fig}, we present the F1-Score for Sao Paolo, Manhattan, Teruel, and Valencia, respectively. A consistent pattern emerges across these figures, with 100\% F1-Score achieved through Centralized Training, dropping to 98\% with Federated Learning (decentralized training), and rebounding to 99.67\% with the application of SMOTE in Manhattan. The results for Sao Paolo follow a similar trend. In Teruel, we started with a 100\% F1-Score using XGBoost and training with 70\% of each dataset, shifted to 98.03\% with Federated Learning and local training, and further increased to 99.81\% with SMOTE. The most notable increase is observed in Valencia, where we traded 100\% F1-Score for enhanced privacy, resulting in a 97.02\% F1-Score with local training. By applying the threshold and SMOTE, we recovered to 99.78\% F1-Score in Valencia.

\begin{figure}
    \centering
	\subfloat[Manhattan]{\includegraphics[width=0.5\linewidth]{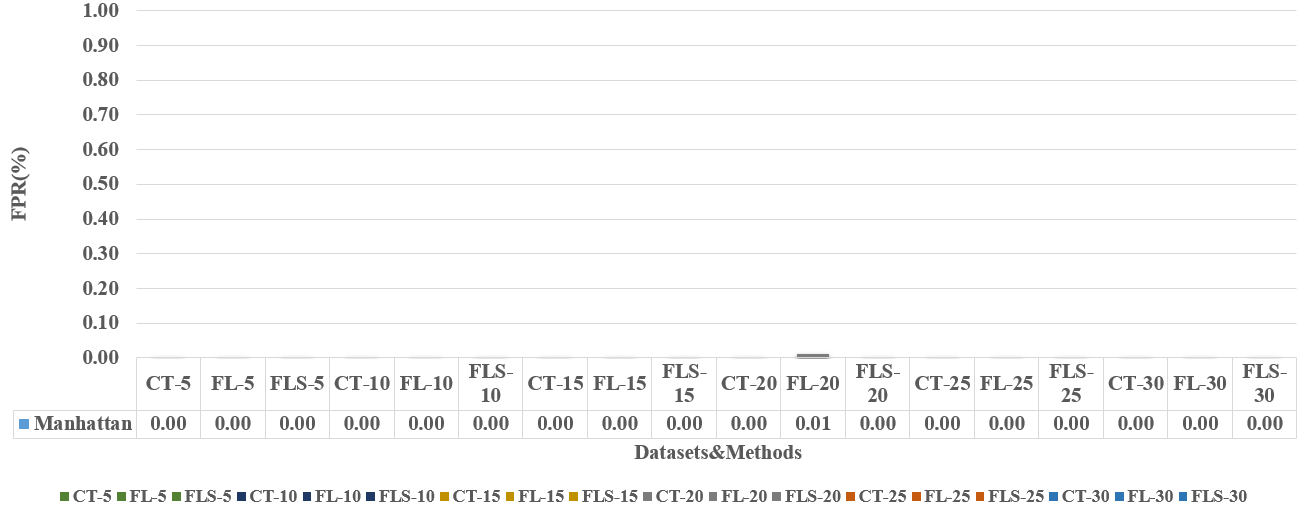}\label{ADS_M_FPR_fig}}
	\subfloat[Sao Paolo]{\includegraphics[width=0.5\linewidth]{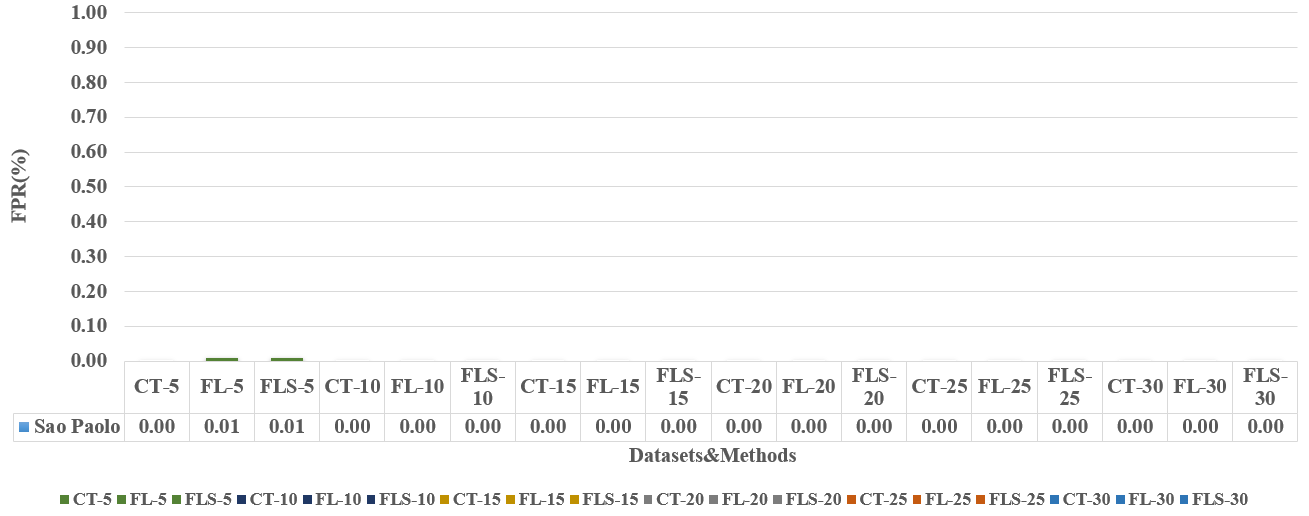}\label{ADS_S_FPR_fig}} \\
	\subfloat[Teruel]{\includegraphics[width=0.5\linewidth]{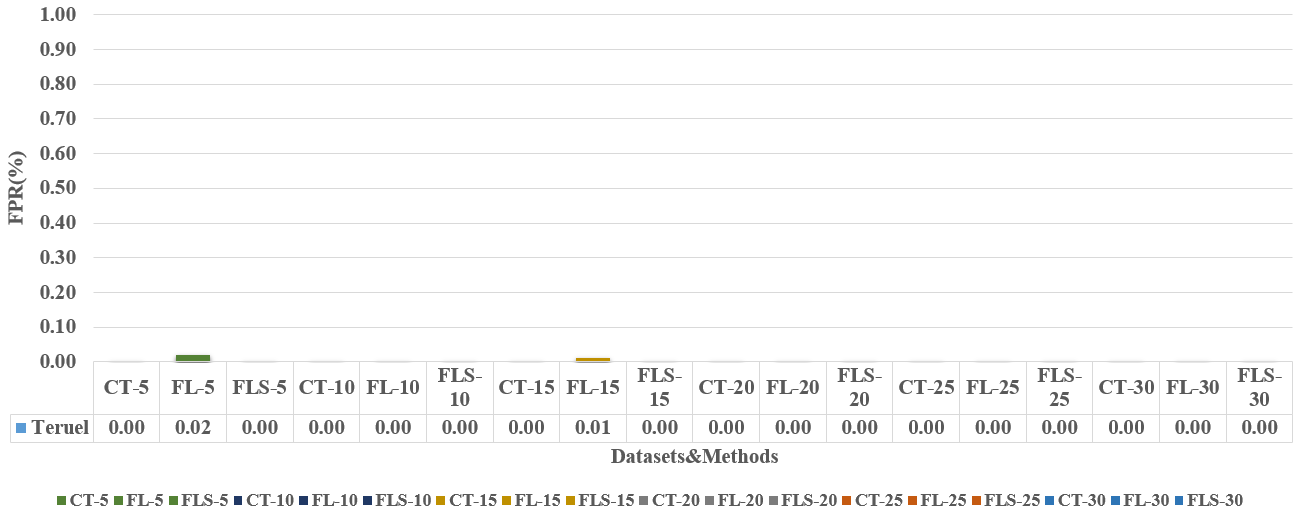}\label{ADS_T_FPR_fig}}
        \subfloat[Valencia]{\includegraphics[width=0.5\linewidth]{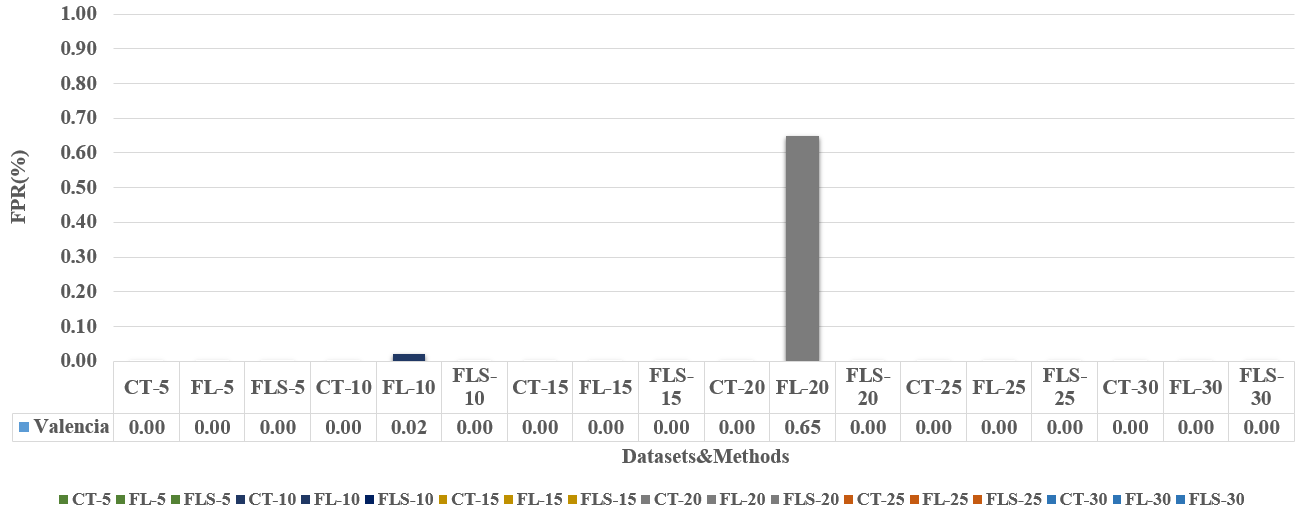}\label{ADS_V_FPR_fig}}
    \caption{FLS-Attack Onset Detection-F1-Score}
    \label{fig:FLS-FPR}
\end{figure}

In Figures \ref{ADS_M_FPR_fig} to \ref{ADS_V_FPR_fig}, we observe a solitary case of the FPR in Valencia datasets, where the value is just under 1\%. In all other cases, the FPR remains at 0 or under 0.01\%, indicating exceptionally good performance. The consistently low FPRs across different scenarios and datasets emphasize the robustness and efficacy of ADVENT in minimizing false positives.

While the FPR consistently yielded favorable results across different scenarios tested on various datasets, challenges surfaced in the FNR. The scarcity of malicious samples during the local training phase significantly impacts the FNR, consequently influencing the DR and F1-Score. However, implementing a threshold, defined based on results from diverse datasets and cities, leads to notable improvements.

\begin{figure}
    \centering
	\subfloat[Manhattan]{\includegraphics[width=0.5\linewidth]{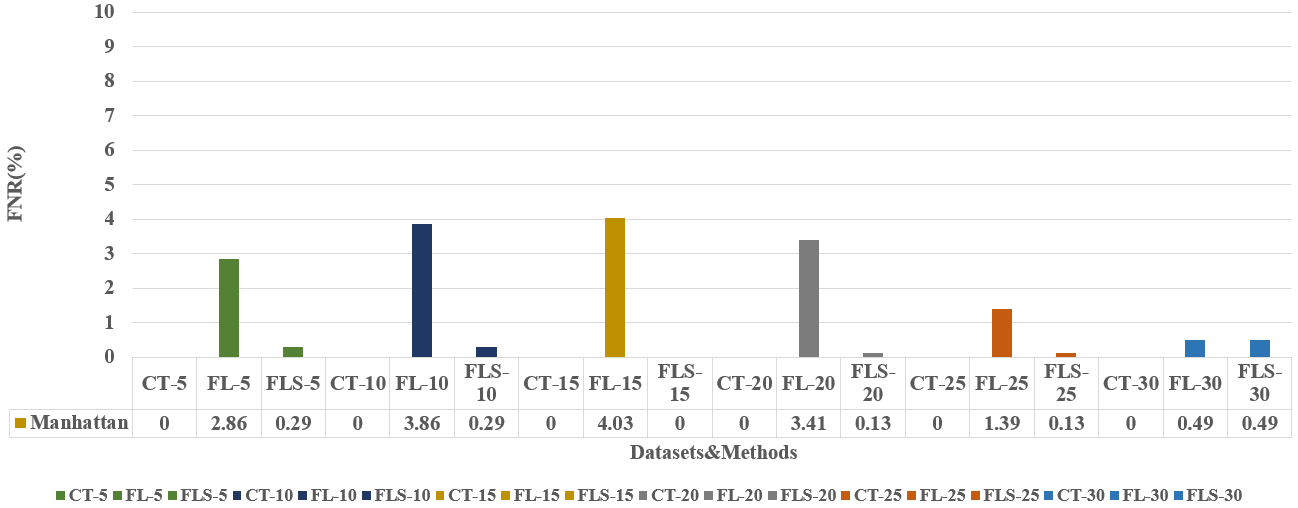}\label{ADS_M_FNR_fig}}
	\subfloat[Sao Paolo]{\includegraphics[width=0.5\linewidth]{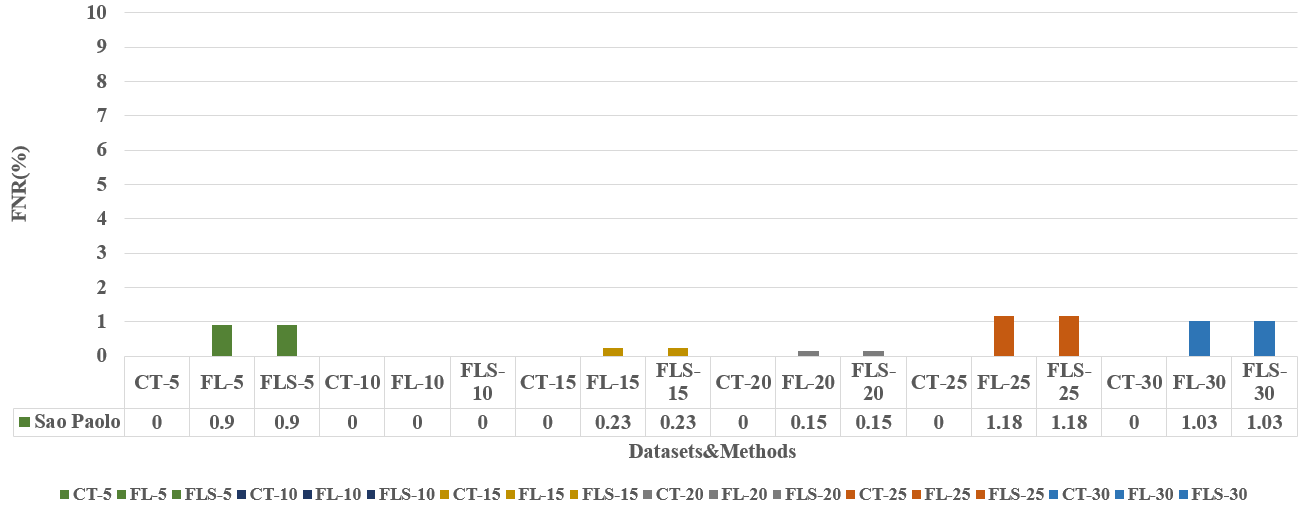}\label{ADS_S_FNR_fig}} \\
	\subfloat[Teruel]{\includegraphics[width=0.5\linewidth]{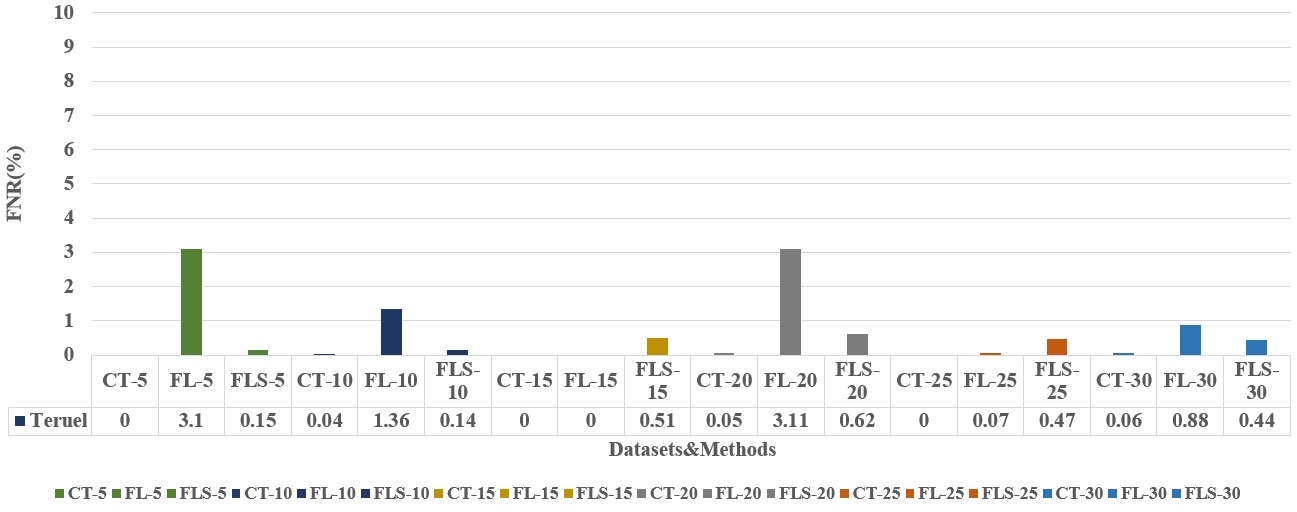}\label{ADS_T_FNR_fig}}
        \subfloat[Valencia]{\includegraphics[width=0.5\linewidth]{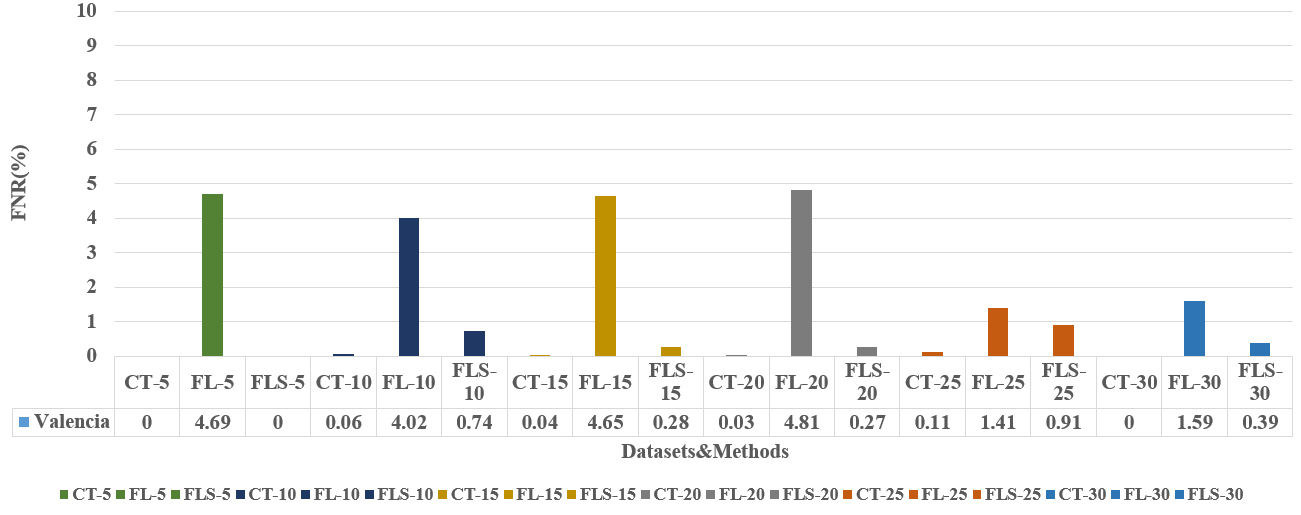}\label{ADS_V_FNR_fig}}
    \caption{FLS-Attack Onset Detection-F1-Score}
    \label{fig:FLS-FNR}
\end{figure}

In Manhattan datasets, where the CT achieved a 0\% FNR, local training resulted in a 2.67\% FNR, which further decreased to 0.22\% with the application of SMOTE. In Sao Paolo, which served as the baseline for threshold determination, no substantial change was observed. Teruel exhibited fewer changes compared to Manhattan and Valencia, moving from a 0.03\% FNR to 1.42\% during local training, and then reverting to 0.39\%. Valencia, following a similar pattern to previous results, witnessed significant fluctuations. Starting with a 0.04\% FNR in our baseline method, it increased to 3.53\% with FedXGBllr and decreased to 0.43\% with the application of SMOTE. The results are shonw in Figures \ref{ADS_M_FNR_fig} to \ref{ADS_V_FNR_fig}.

\begin{figure}[!t]
\centering
\includegraphics[width=4.7in]{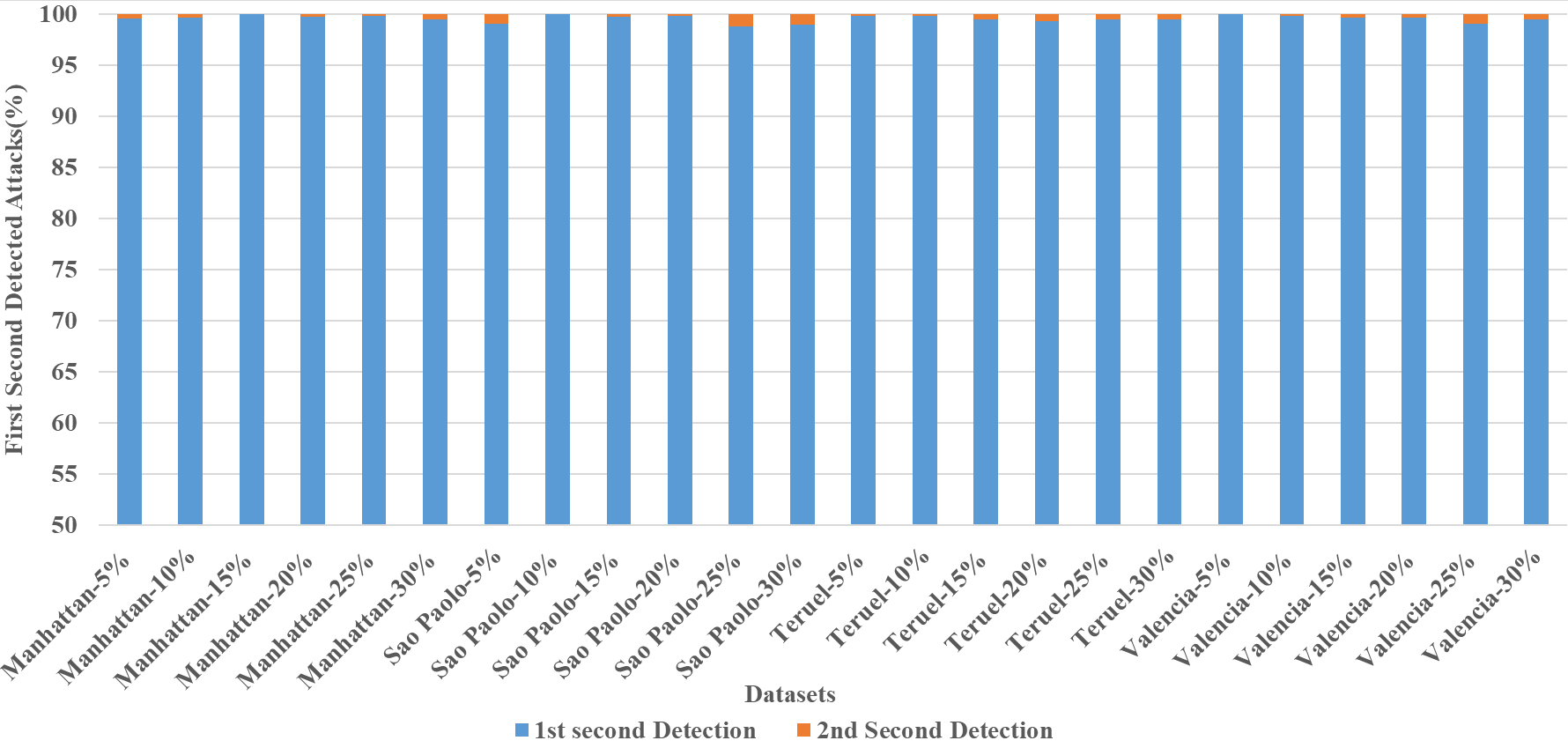}
\caption{Valencia-FLS-Attack Onset Detection-FPR}
\label{First_fig}
\end{figure}

The final factor we explore is how quickly attacks are detected to provide insight into the effectiveness of ADVENT in preventing further harm to vehicles and the network in realtime. Figure \ref{First_fig} illustrates the results, indicating an average of 99.6\% of all attacks being detected within the first second of commencement. This showcases the high efficacy of our method in swiftly identifying and responding to attacks as they unfold, contributing to the overall security and resilience of the vehicular ad-hoc network.

\subsubsection*{\bf Results of Malicious Node Detection}: Malicious Node Detection involves the utilization of the Median Absolute Deviation (MAD) technique. Subsequently, nodes share the lists of suspicious vehicles with a central server, enabling the server to compile a finalized list of malicious vehicles, which is then distributed to all vehicles. Following these steps, we observed significant improvements, particularly in terms of DR and FNR. These enhancements were particularly noteworthy when compared to our previous work, as discussed in \cite{ours2}, where we encountered a higher FNR due to variations in vehicle experience and the packets received.
It's worth noting that on the server side, ADVENT utilizes a frequency-based threshold to mitigate the FPR. This threshold enables ADVENT to reduce the impact of false reports. We conducted an experiment with a threshold value set to 2 in this study to evaluate its effect compared to not applying the threshold. It's essential to mention that the threshold value can be adjusted based on the specific network conditions and features, allowing for adaptability to different scenarios.

\begin{figure}
    \centering
	\subfloat[Manhattan]{\includegraphics[width=0.5\linewidth]{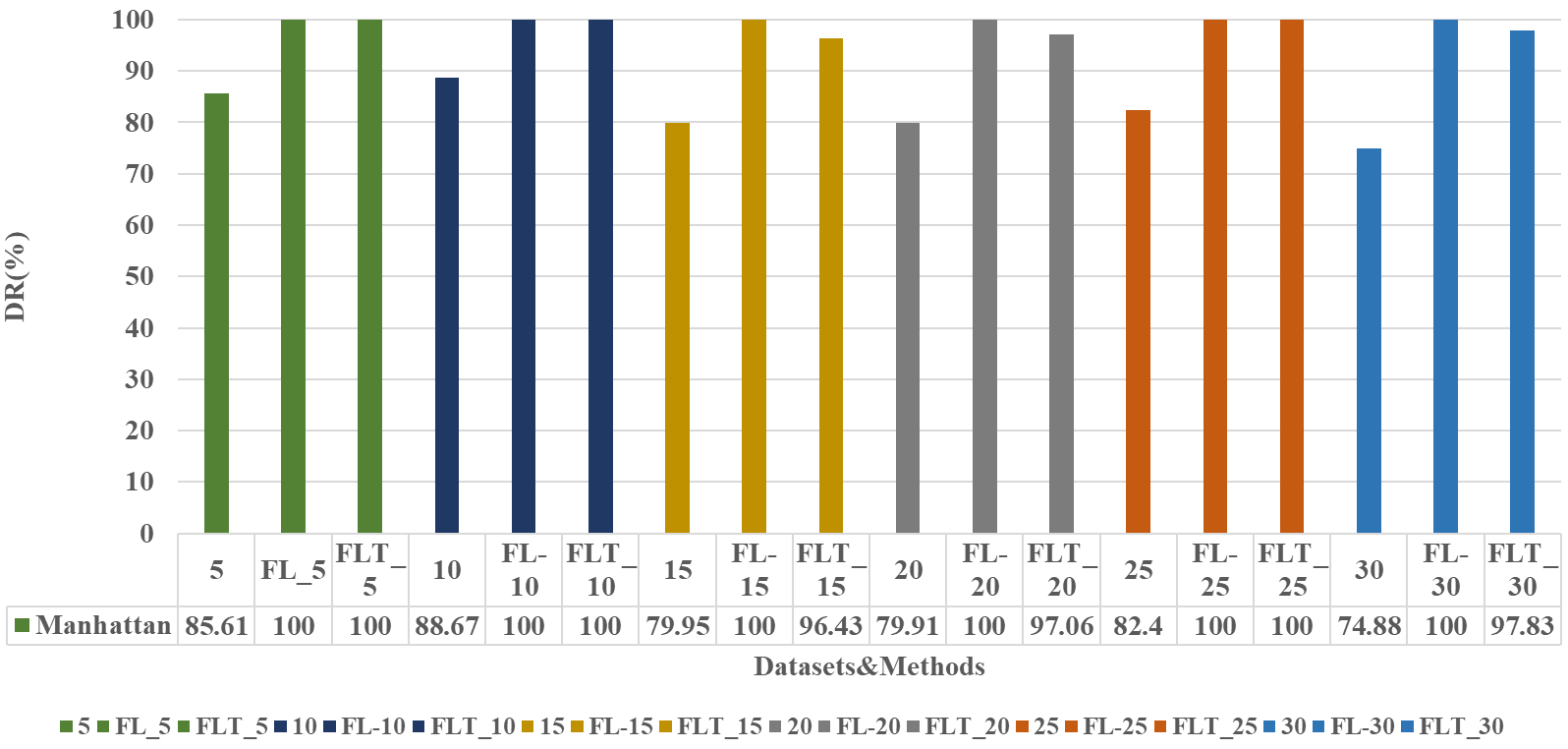}\label{MD_DR_M_fig}}
	\subfloat[Sao Paolo]{\includegraphics[width=0.5\linewidth]{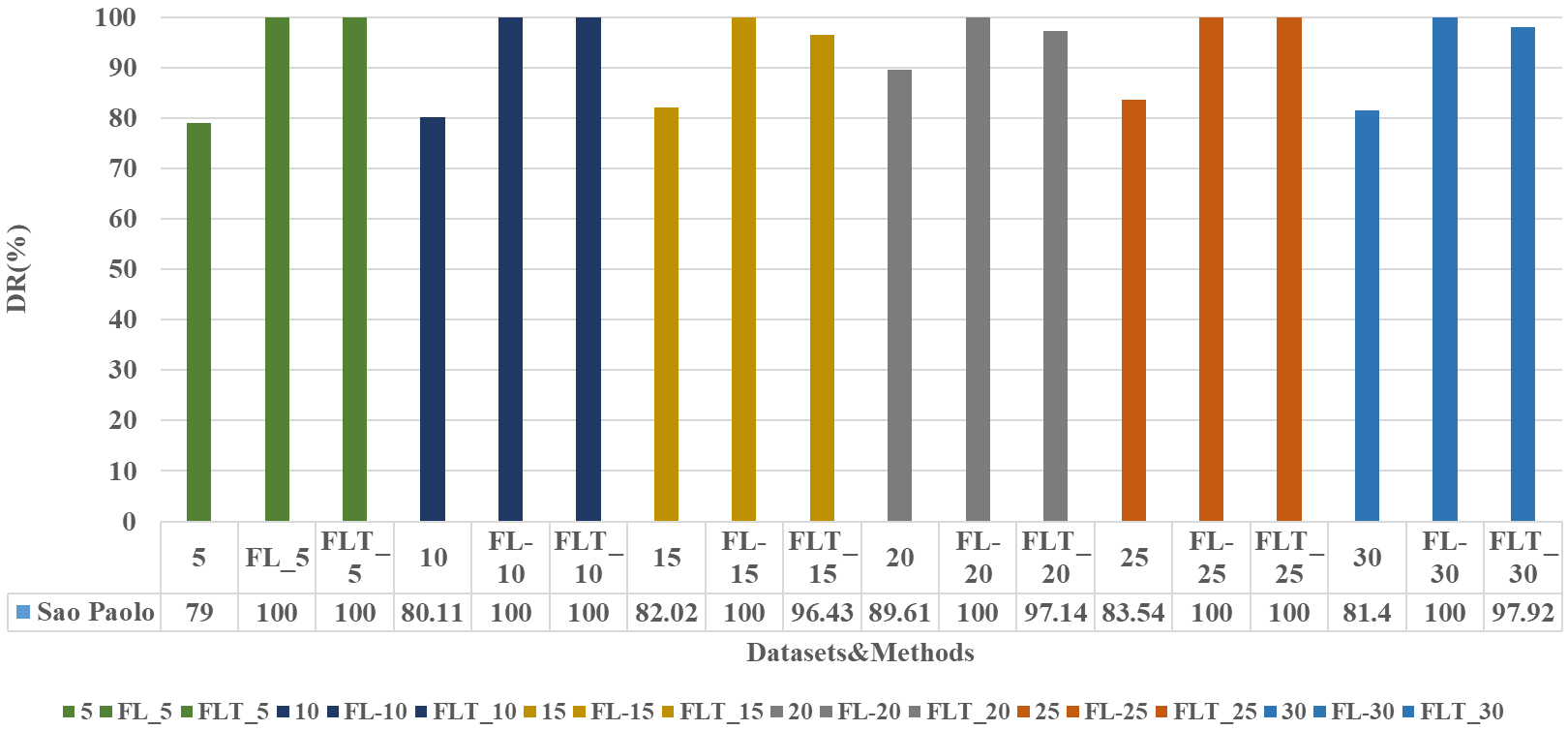}\label{MD_DR_S_fig}} \\
	\subfloat[Teruel]{\includegraphics[width=0.5\linewidth]{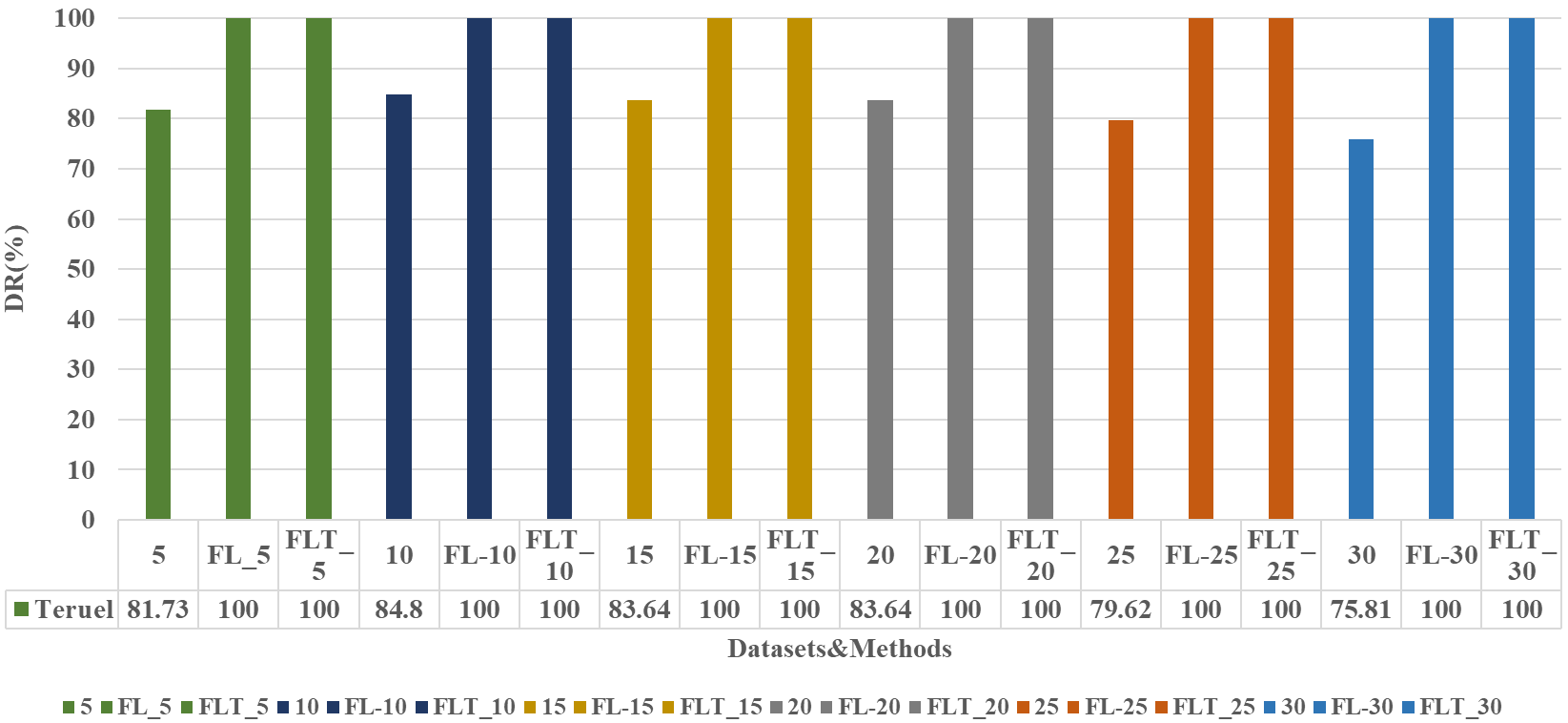}\label{MD_DR_T_fig}}
        \subfloat[Valencia]{\includegraphics[width=0.5\linewidth]{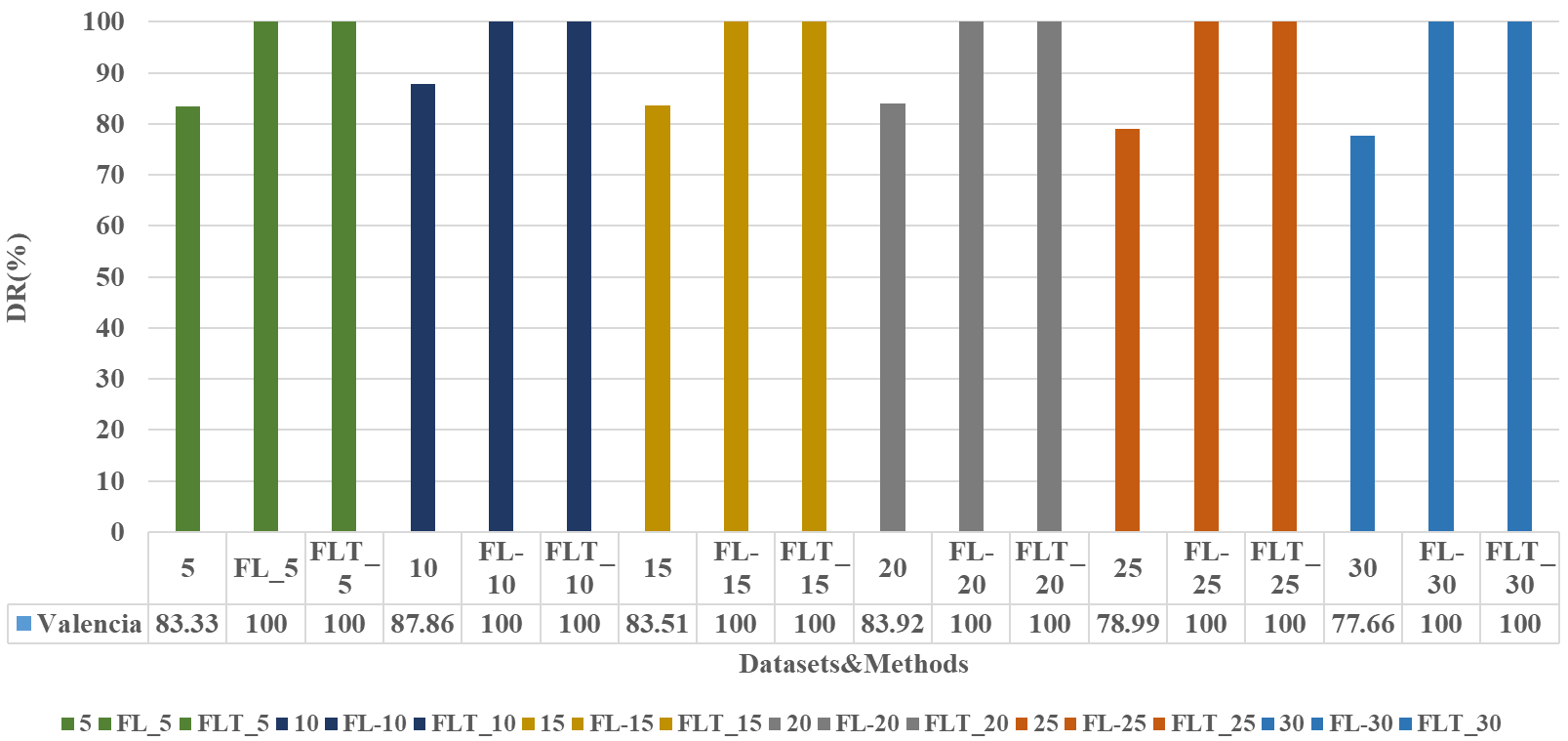}\label{MD_DR_V_fig}}
    \caption{Malicious Node Detection-Detection Rate}
    \label{fig:MD-DR}
\end{figure}

Through the utilization of this vehicle-specific testing methodology and the computation of average metrics, we obtain valuable insights into the performance of the proposed method on individual vehicles. In Fig \ref{fig:MD-DR}, we present the DR results for the three compared methods. The first method, as outlined in \cite{ours}, applies the MAD technique on the same 24 datasets and computes the average results from all nodes. The second method involves using Federated Learning with the sharing of suspected vehicle lists with the server to obtain the malicious node list. The third method explores the impact of setting a threshold value to 2, designating a vehicle as an attacker if it is reported by two or more vehicles.

Notably, the application of the Federated Learning method yields a remarkable improvement in the DR, reaching 100\% for all cities. This marks an average increase of 17.45\% for Valencia, 18.46\% for Teruel, 18.10\% for Manhattan, and 17.39\% for datasets pertaining to Sao Paulo. However, the utilization of the threshold does lead to a decrease in the average DR for Manhattan and Sao Paulo, dropping it to 98.55\%. On the other hand, Teruel and Valencia maintain a 100\% DR even after the application of the threshold.\ref{MD_DR_M_fig} to \ref{MD_DR_V_fig}
Additionally, a pivotal achievement of this method is the notable reduction in disparities among results from various cities, especially when we use pure federated learning aggregation method, where all the cities acheive 100\% DR. This shows the effectiveness of integrating these steps and methodologies in mitigating the influence of geographical patterns and attackers' distribution, both of which were discussed in previous work \cite{ours, ours2}. The significance of this achievement is rooted in the understanding that geographical variations can profoundly impact the efficacy of each method, and the proposed approach successfully addresses these challenges.

\begin{figure}
    \centering
	\subfloat[Manhattan]{\includegraphics[width=0.5\linewidth]{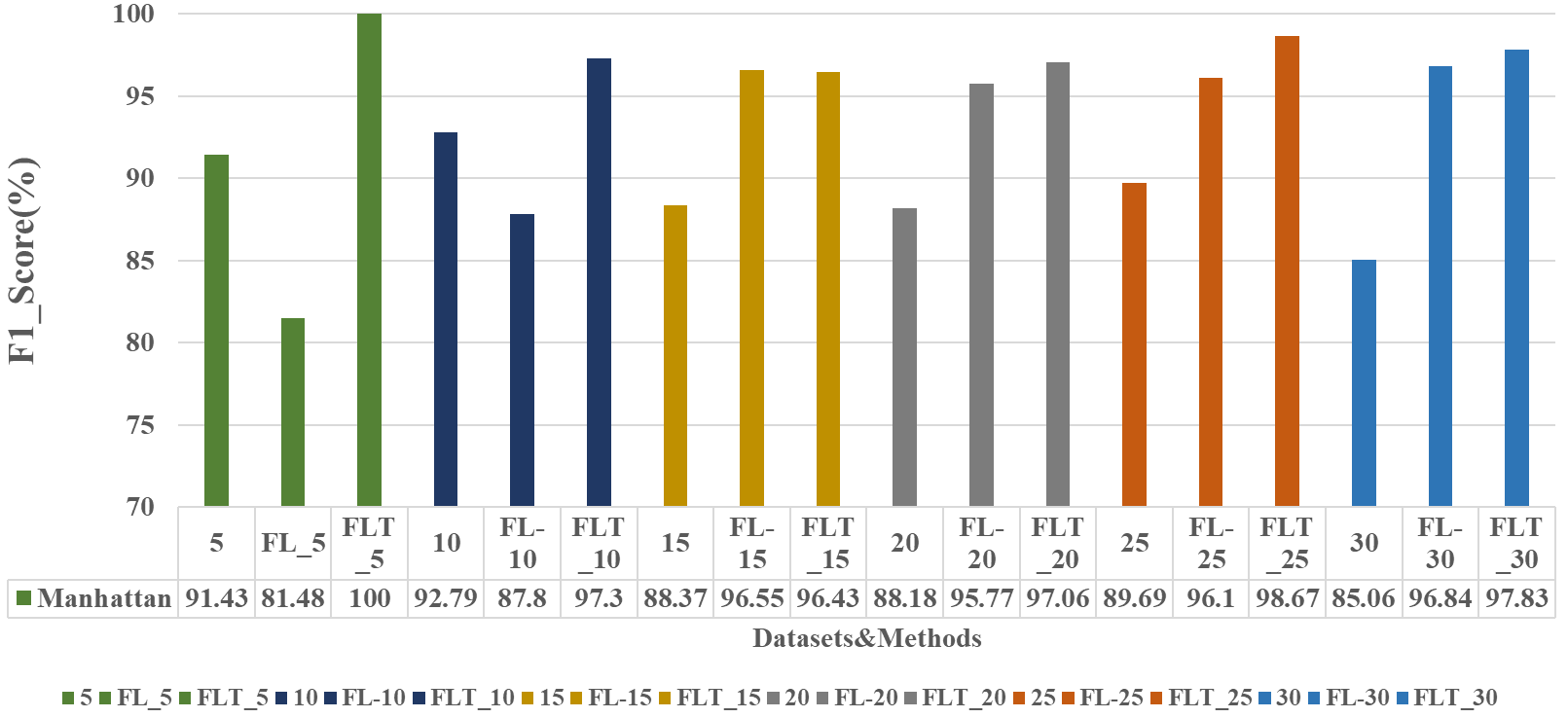}\label{MD_F1_M_fig}}
	\subfloat[Sao Paolo]{\includegraphics[width=0.5\linewidth]{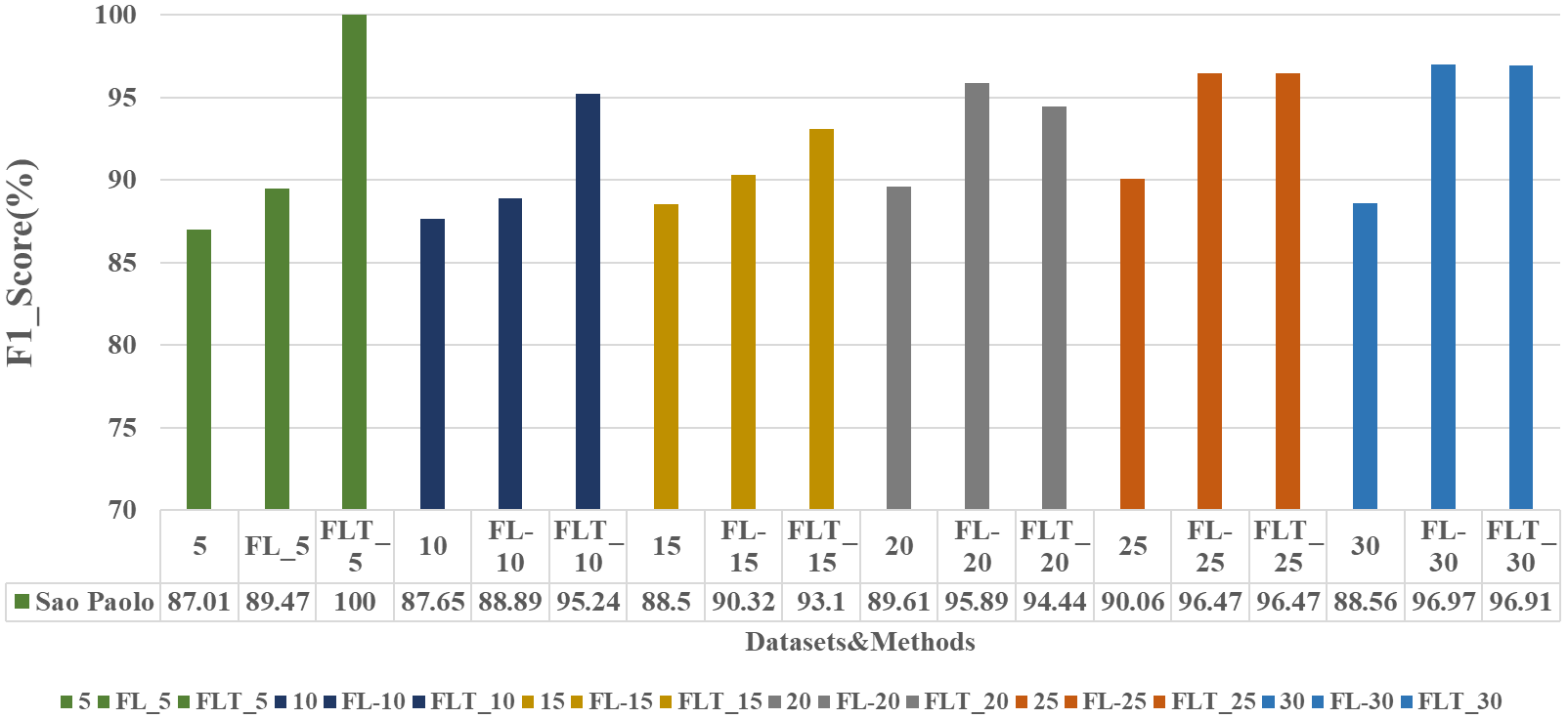}\label{MD_F1_S_fig}} \\
	\subfloat[Teruel]{\includegraphics[width=0.5\linewidth]{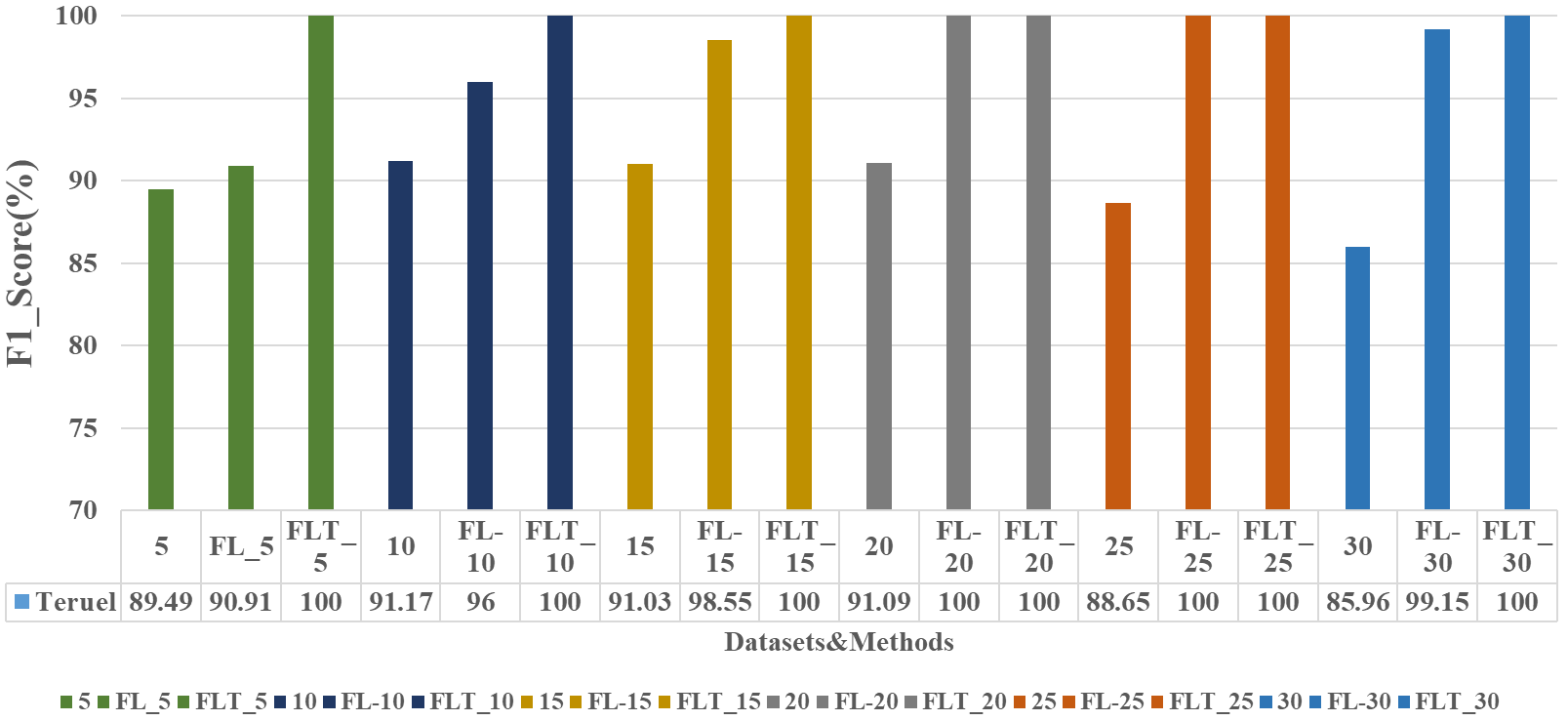}\label{MD_F1_T_fig}}
        \subfloat[Valencia]{\includegraphics[width=0.5\linewidth]{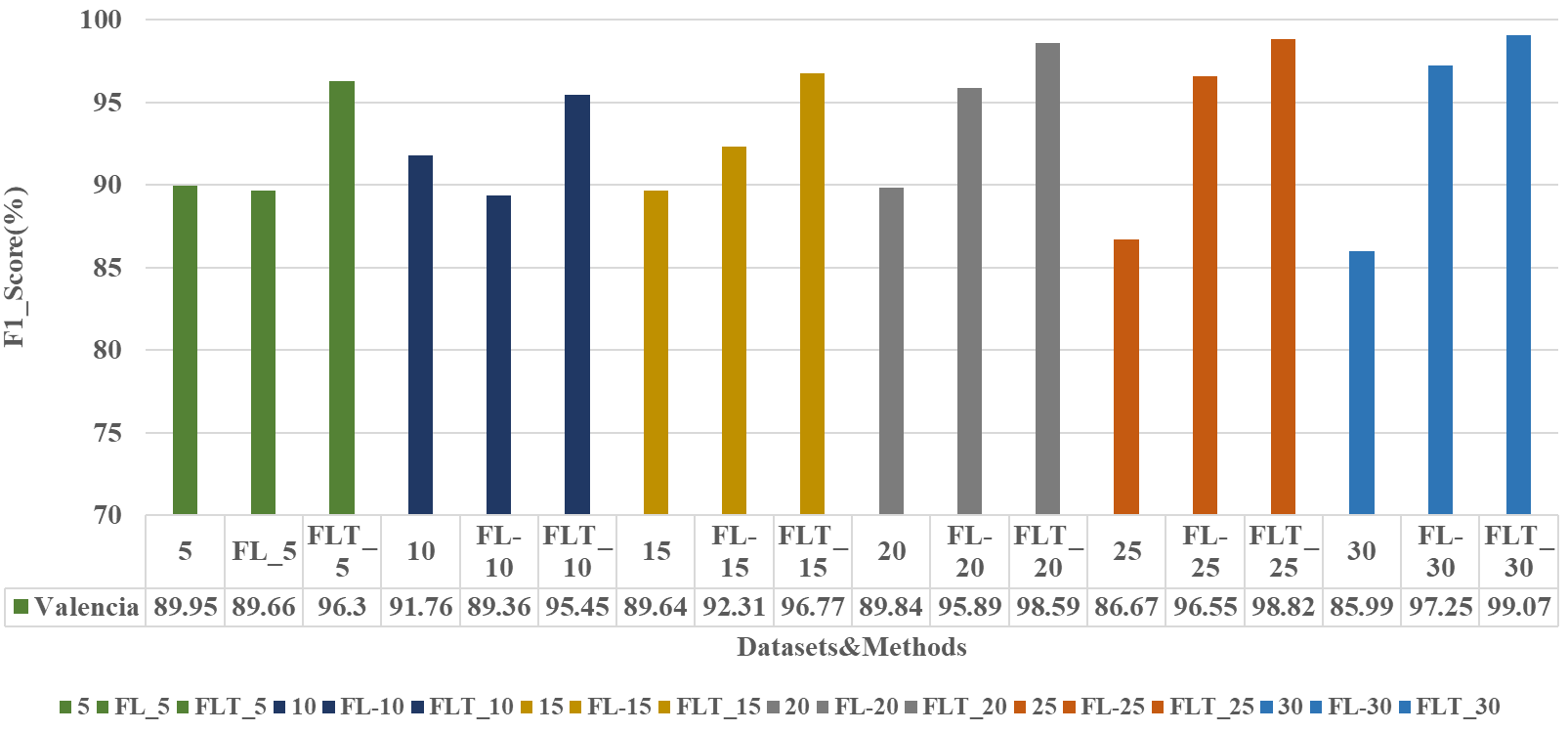}\label{MD_F1_V_fig}}
    \caption{Malicious Node Detection-F1-Score}
    \label{fig:MD-F1}
\end{figure}

The next metric under discussion is the F1 score, and Figures \ref{MD_F1_M_fig} to \ref{MD_F1_V_fig} provide a detailed breakdown of the results across all 24 datasets. On average, for Sao Paulo, the F1 score increases from 88.56\% with the MAD techniques to 93\% with the Federated Learning and further improves to 96.03\% with the application of the threshold. Similarly, Manhattan, which initially averages an F1 score of 89.25\% with MAD, reaches 92.43\% when using Federated Learning and aggregating the lists, ultimately achieving an impressive 97.88\% F1 score with the threshold.

In the Teruel datasets, even better results are obtained, starting at 89.57\% F1 score and moving to 97.43\% for the Federated Learning and ultimately reaching 100\% with the integration of the threshold. Valencia exhibits a similar pattern to Sao Paulo and Manhattan, with the F1 score increasing from 88.97\% to 93.50\%, reflecting a 4.53\% improvement via Federated Learning, and then a 97.50\% F1 score with an additional 4\% improvement, upon applying the threshold.

\begin{figure}
    \centering
	\subfloat[Manhattan]{\includegraphics[width=0.5\linewidth]{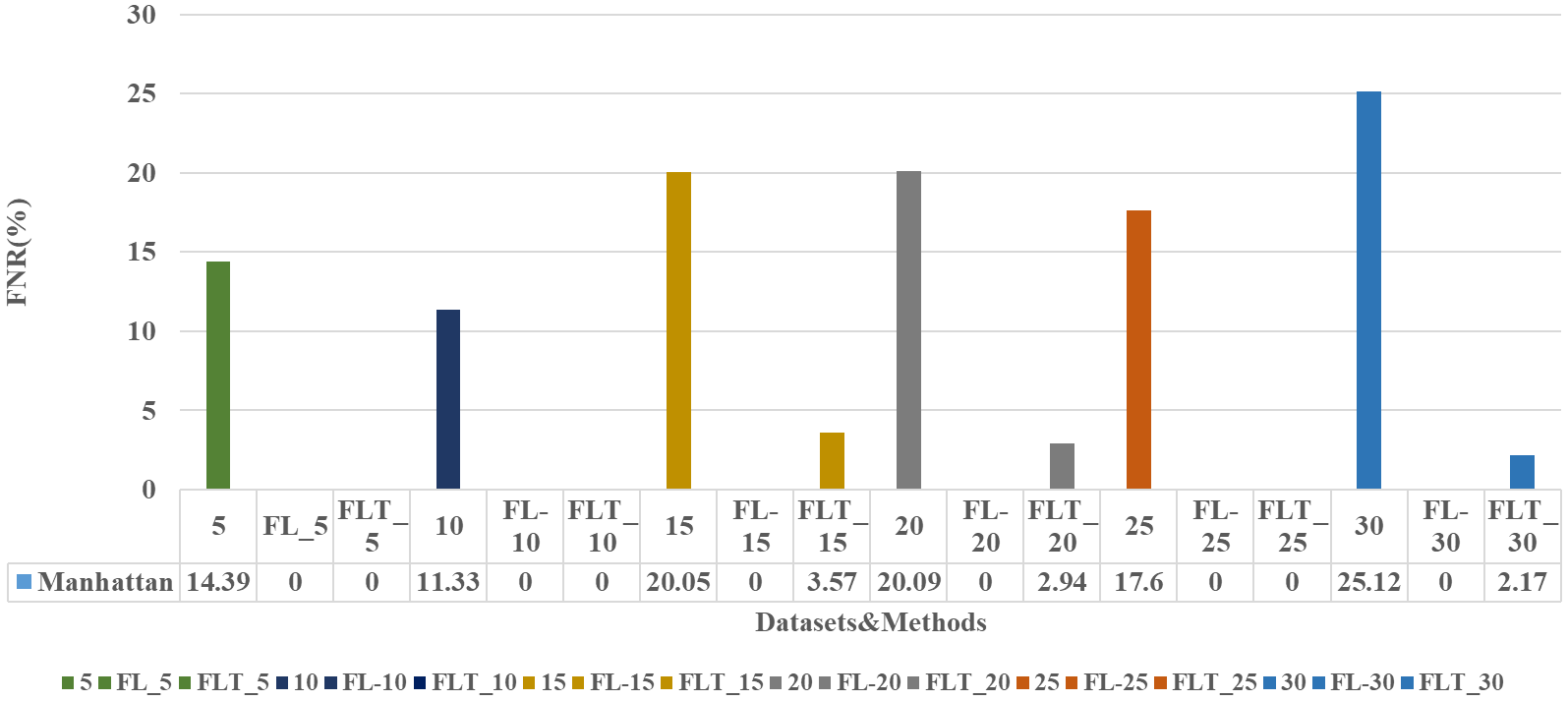}\label{MD_FNR_M_fig}}
	\subfloat[Sao Paolo]{\includegraphics[width=0.5\linewidth]{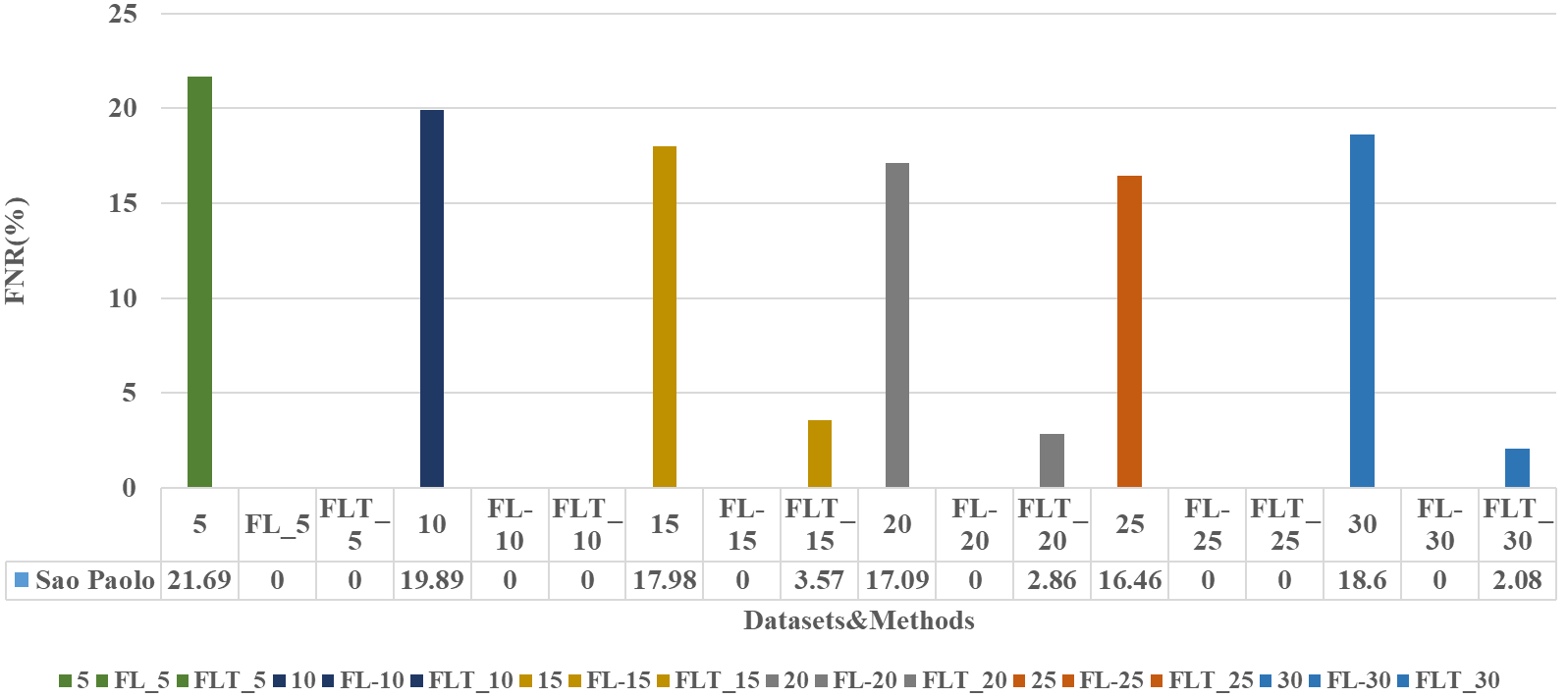}\label{MD_FNR_S_fig}} \\
	\subfloat[Teruel]{\includegraphics[width=0.5\linewidth]{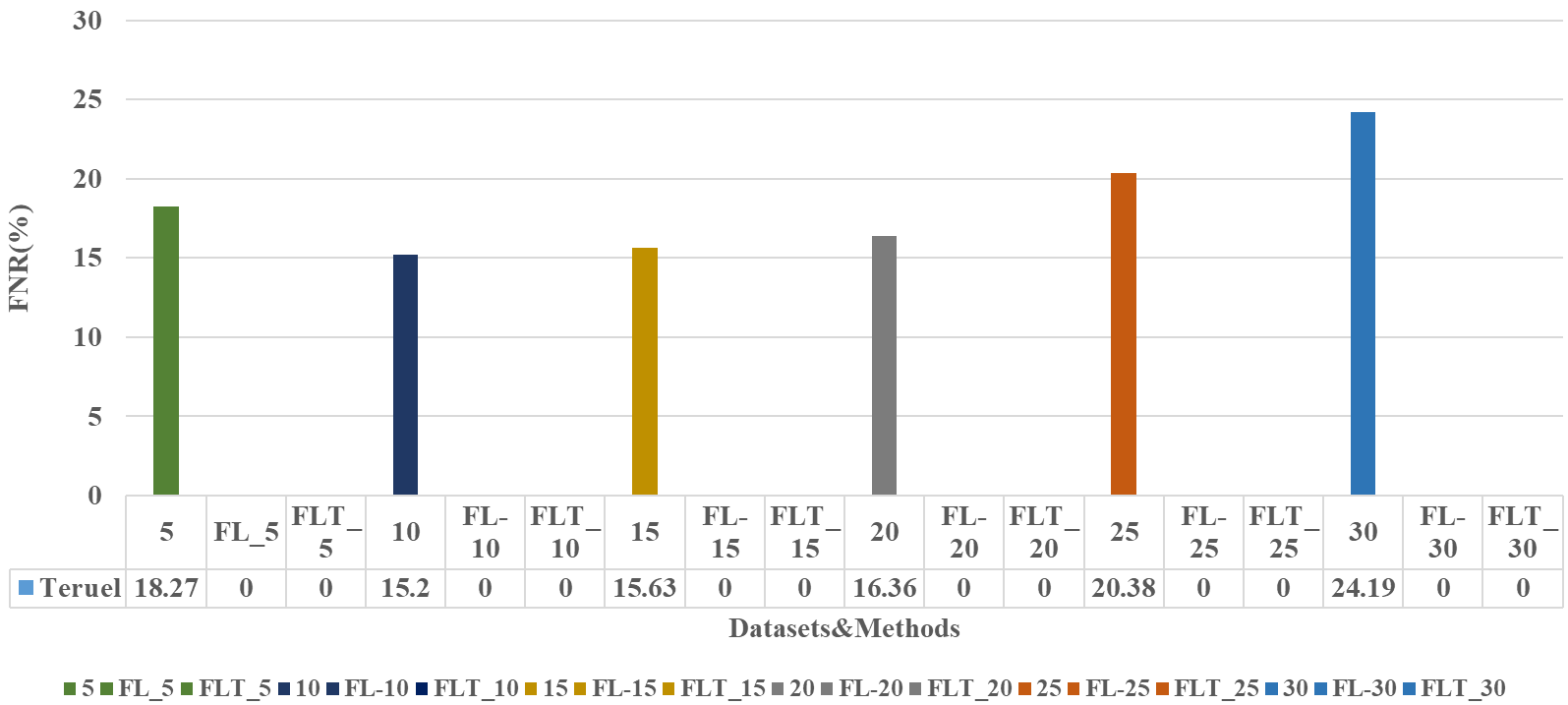}\label{MD_FNR_T_fig}}
        \subfloat[Valencia]{\includegraphics[width=0.5\linewidth]{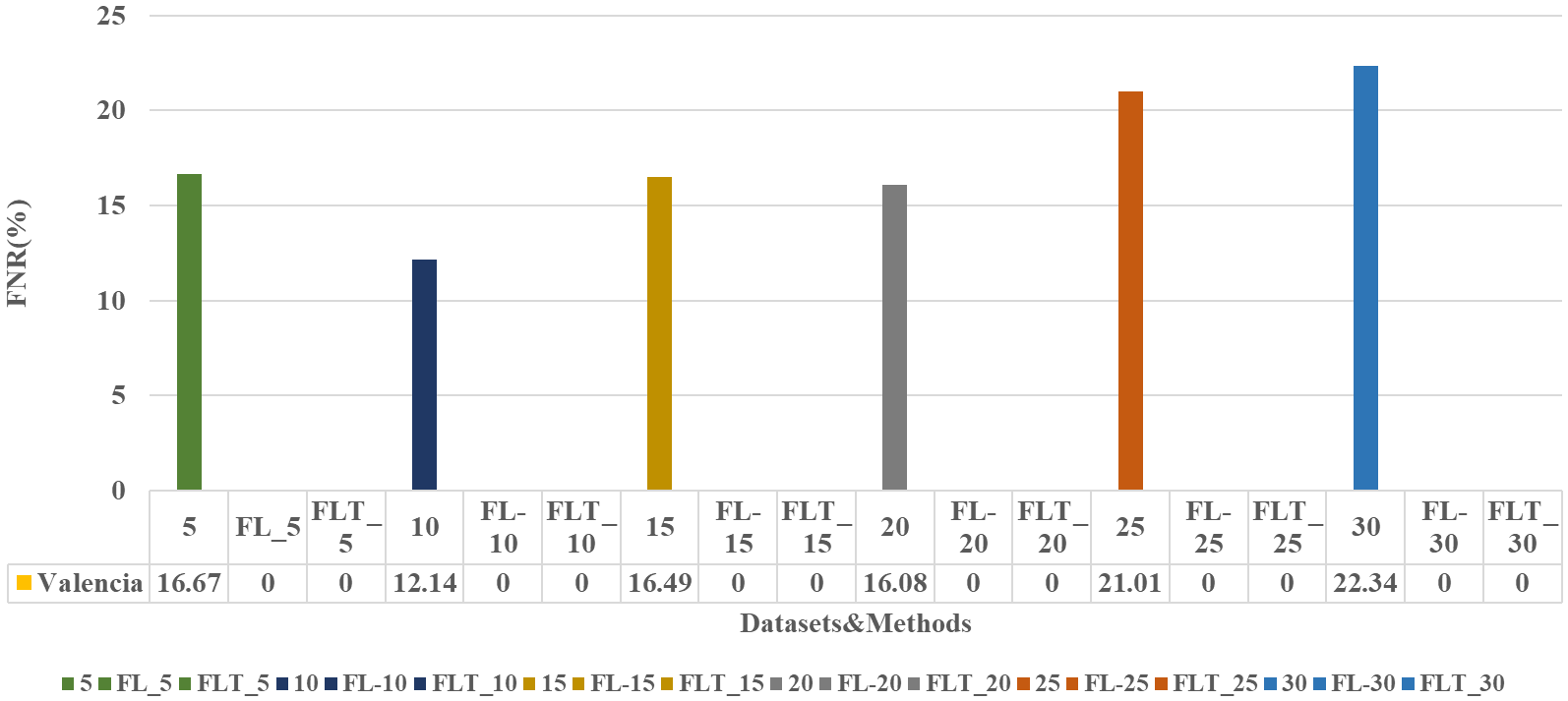}\label{MD_FNR_V_fig}}
    \caption{Malicious Node Detection-False Negative Rate}
    \label{fig:MD-FNR}
\end{figure}

In Figures \ref{MD_FNR_M_fig} to \ref{MD_FNR_M_fig}, the results for the FNR are shown, and the outcomes are quite remarkable. We observed a substantial decrease in the FNR for all cities, namely Sao Paulo, Manhattan, Teruel, and Valencia. The FNR reduced from 18.62\%, 18.10\%, 18.34\%, and 17.45\%, respectively, to around 0\% when applying Federated Learning, benefiting from the collective wisdom of all nodes and aggregating their experiences. Furthermore, after setting the threshold to 2, we still experienced some FNR in Sao Paulo and Manhattan, averaging at 1.42\% and 1.45\%, respectively. However, it is important to note that in the other two cities, Teruel and Valencia, the FNR remained at 0. Nonetheless, when compared to the previous work in \cite{ours2}, this study demonstrates a significant improvement, with a meaningful reduction in FNR.

\begin{figure}
    \centering
	\subfloat[Manhattan]{\includegraphics[width=0.5\linewidth]{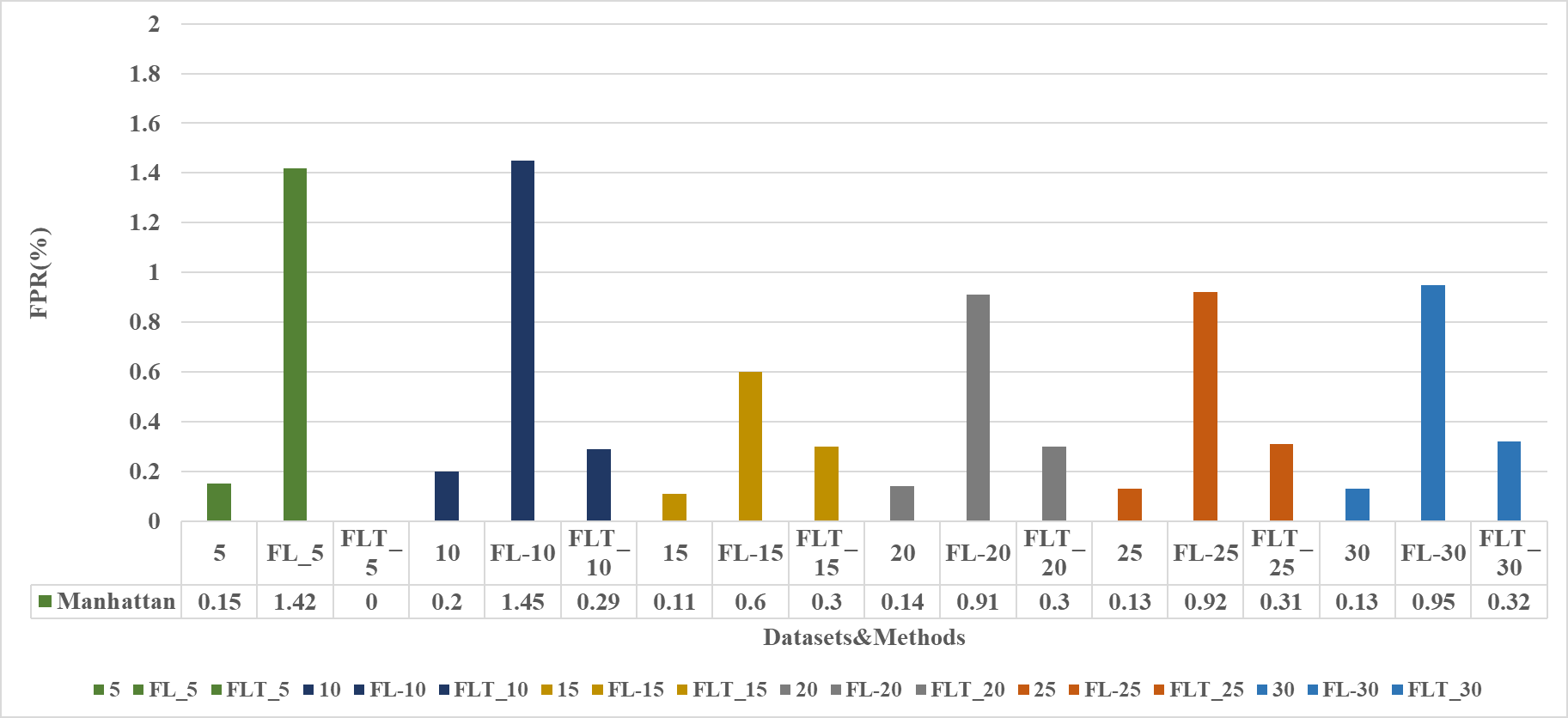}\label{MD_FPR_M_fig}}
	\subfloat[Sao Paolo]{\includegraphics[width=0.5\linewidth]{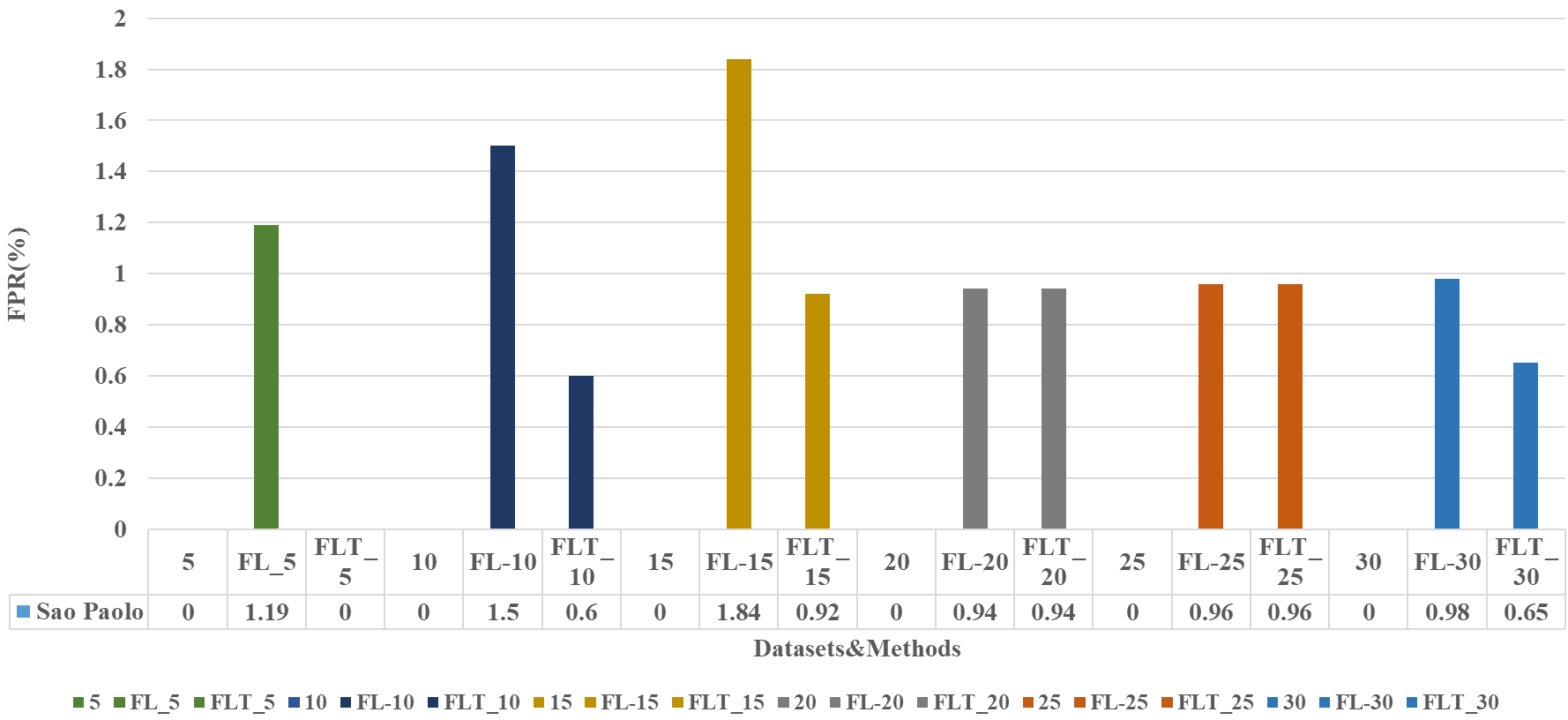}\label{MD_FPR_S_fig}} \\
	\subfloat[Teruel]{\includegraphics[width=0.5\linewidth]{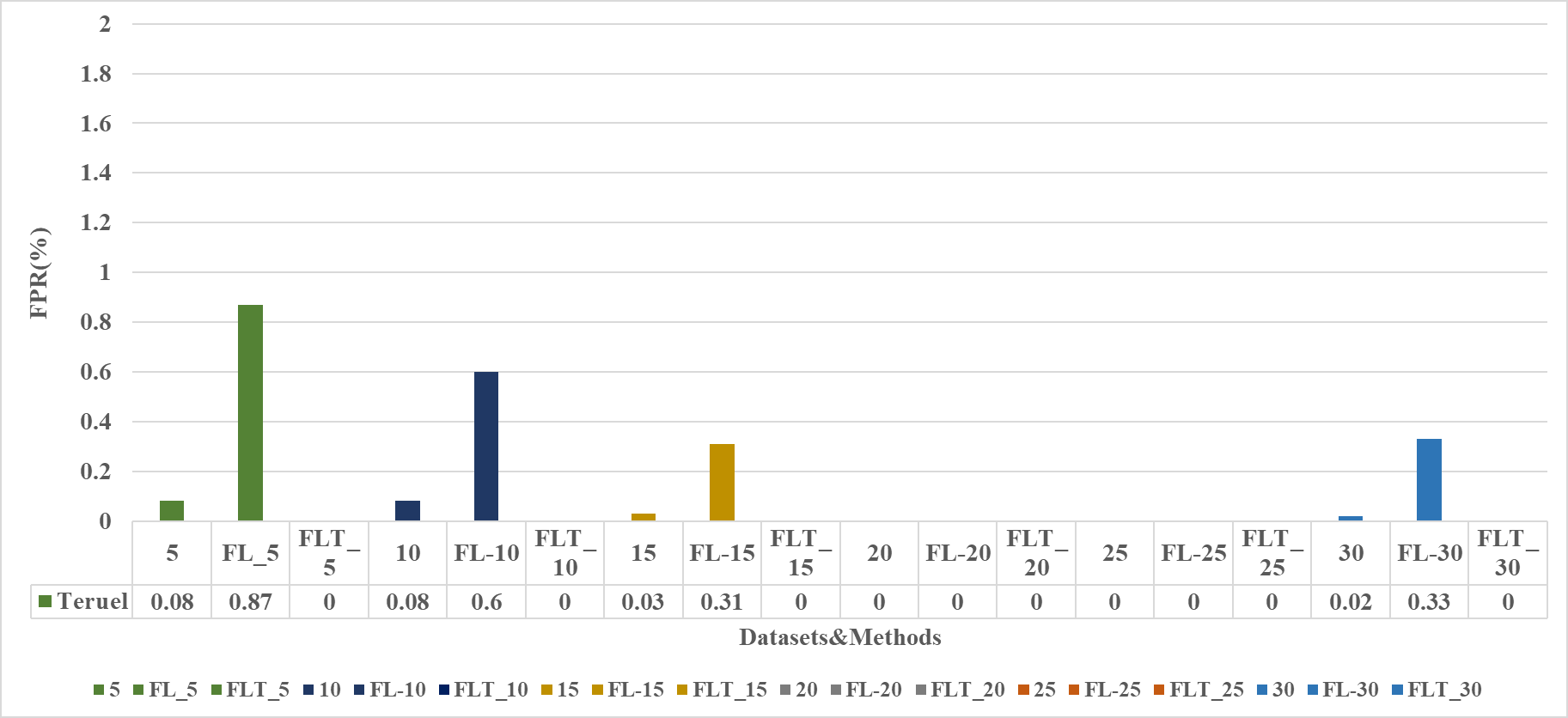}\label{MD_FPR_T_fig}}
        \subfloat[Valencia]{\includegraphics[width=0.5\linewidth]{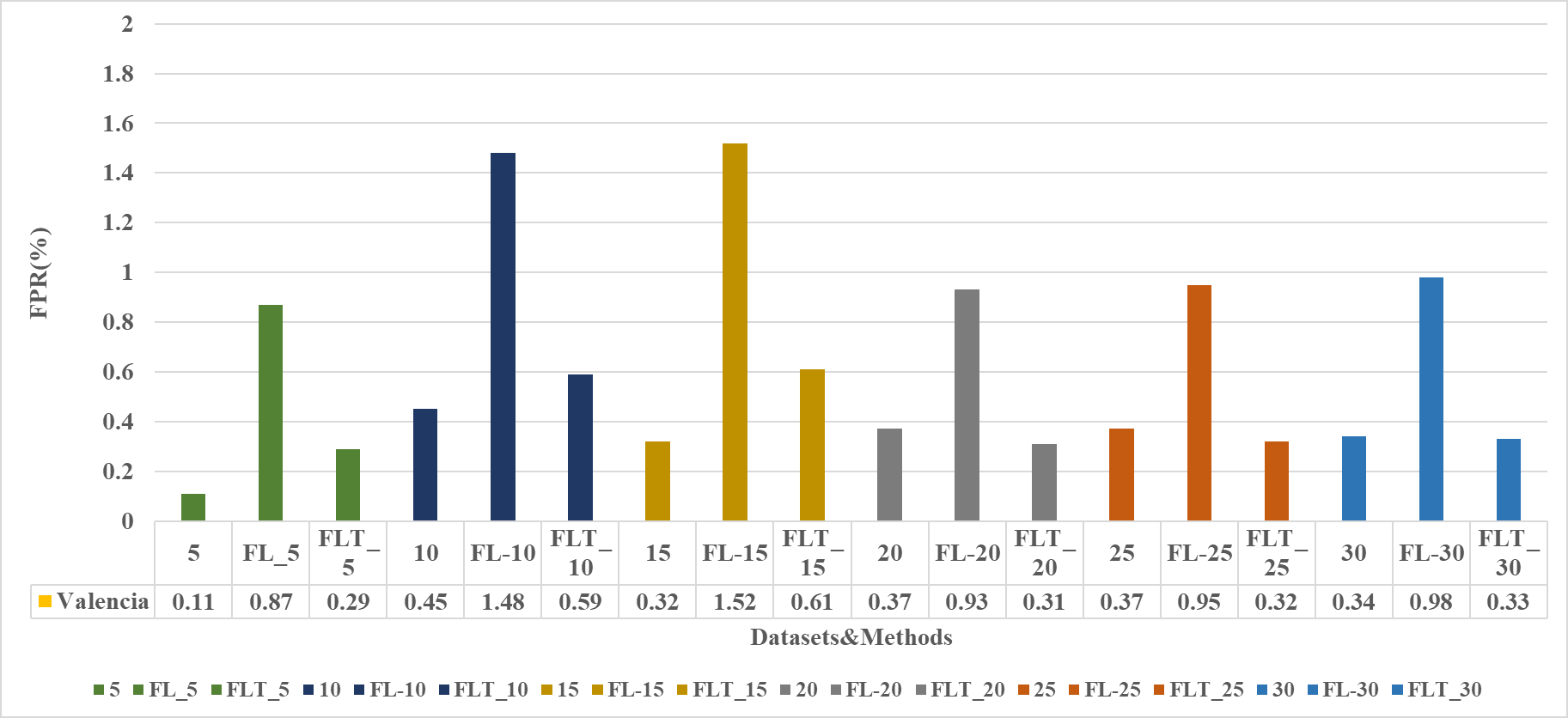}\label{MD_FPR_V_fig}}
    \caption{Malicious Node Detection-False Positive Rate}
    \label{fig:MD-FPR}
\end{figure}

In Fig. \ref{fig:MD-FPR}, we present the results for the FPR, highlighting the effects of incorporating a threshold and combining it with the Federated Learning method. This approach strikes a balance between FPR and FNR. Initially, with the basic MAD technique, the FPR was very low, measuring at 0.00\%, 0.14\%, 0.03\%, and 0.33\% for Sao Paulo, Manhattan, Teruel, and Valencia, respectively. However, when we apply the Federated Learning method, there is an increase in FPR to 1.23\%, 1.04\%, 0.35\%, and 1.12\% for the same cities. While these FPR values are not very high, applying the threshold helps reduce them to an average of 0.68\% for Sao Paulo, 0.25\% for Manhattan, 0.00\% for Teruel, and 0.41\% for Valencia, resulting in a favorable balance for all four cities.

\subsection{Time complexity}
Table \ref{tab:time} provides statistical insights into the system's real-world performance, presenting the preprocessing time for each dataset in the first column, the time taken to detect the attack onset in the second column, and the duration ADVENT system takes to identify the list of malicious nodes in the third column. According to the statistics in this figure, ADVENT achieves synchronization and generates a unified model for network use within an average of 6 seconds. Moreover, it demonstrates the capability to detect the onset of an attack in the very first second in 99.6\% of the scenarios and identifies malicious behavior in 1.97 seconds in average.

\section{Conclusions and Future Works}
\label{sec:con}
In conclusion, our exploration into the realm of VANETs has uncovered both the potential and the challenges associated with this dynamic and critical communication infrastructure. While VANETs offer real-time solutions for addressing delayed driver responses and enhancing traffic safety, their intrinsic vulnerabilities, such as the ever-changing network topologies and the diverse nature of V2X communications, demand a dedicated focus on security \cite{garip}.

To bolster VANET security, we introduce an innovative 3-step real-time malicious behavior detectorcalled ADVENT. ADVENT seamlessly integrates machine learning and statistical methodologies to ensure the effective detection of attack onsets. The results consistently showcase an impressive average detection rate of 99.6\% for DDoS attack onsets occurring in the first second. It also excels in identifying malicious nodes, a crucial aspect for maintaining the integrity of the network, achieving a remarkable 99.28\% Detection Rate (DR). Importantly, we prioritize data privacy through the implementation of Federated Learning, avoiding the transmission of raw data. This not only enhances efficiency but also minimizes the False Negative Rate (FNR), resulting in a substantial reduction from 18.13\% to 0.72\% FNR, when compared to prior work \cite{ours,ours2}.

Our pioneering preprocessing method stands out for its effectiveness in understanding communication patterns and potential attack scenarios. This method significantly reduces dataset size and processing time, contributing to a more efficient exploration of network dynamics.

Addressing the challenge of imbalanced data, we identify a threshold and implement the Synthetic Minority Over-sampling Technique (SMOTE) to balance datasets. This yields improved outcomes and minimizes the FNR, showcasing a notable improvement in the DR from 97.95\% to 99.59\% on average and a reduction in the FNR from 2.05\% to 0.41\% on average. These contributions collectively establish a comprehensive multi-layered real-time detection method, privacy-enhanced Federated Learning, and an efficient feature selection process.

The robust evaluation of our proposed method consistently demonstrates its remarkable effectiveness, achieving an overall Detection Rate (DR) of 99.66\%. Leveraging Federated Learning significantly reduces the FNR, thereby enhancing overall model performance. The feature selection process contributes to the efficiency of attack detection models. Realistic VANET simulations ensure the reliability of our datasets and the presented results. Furthermore, the implementation of SMOTE effectively balances imbalanced data, leading to improved results and minimized FNR. These contributions collectively reinforce the security and reliability of VANETs, marking a significant step forward in addressing the evolving challenges in vehicular communication networks.

For future work, an essential avenue involves expanding the system's applicability by testing it on diverse datasets. This includes datasets with varying characteristics and scenarios to assess the system's adaptability and generalization capabilities. Additionally, exploring the system's performance against different types of attacks beyond DDoS will provide valuable insights into its robustness and effectiveness in a broader range of security threats. By systematically evaluating the system's response to various datasets and attack scenarios, we can enhance its versatility and establish a more comprehensive understanding of its capabilities in real-world environments.
Furthermore, there is potential for refining the model-building aspect. The discussed different approaches could be further explored, discussed, and tested to identify the most effective approach. Additionally, there is room to work on optimizing communication links and exploring various choices to enhance the overall efficiency and reliability of the system. These future directions will contribute to the continual improvement and advancement of the proposed system in tackling evolving challenges in vehicular communication networks.

\section*{Acknowledgment}
This research is made possible through the support of the Mitacs funding program. The research is conducted at the Dalhousie Network and Information Management System (NIMS) Lab and the Cybersecurity Lab at New York Institute of Technology in Vancouver, Canada.

\bibliographystyle{IEEEtran}
\bibliography{IoT}

\end{document}